\documentclass[aps,prb,twocolumn,nofootinbib,notitlepage,superscriptaddress,longbibliography,preprintnumbers]{revtex4-2}
\usepackage{times}
\usepackage{graphicx}
\usepackage[whole]{bxcjkjatype} 

\usepackage[top=30truemm,bottom=30truemm,left=25truemm,right=25truemm]{geometry}

\usepackage[colorlinks=true,linktoc=page,citecolor=red,linkcolor=blue]{hyperref}	
\graphicspath{{./figs/}}  
\usepackage[usenames]{xcolor}
\usepackage{amsmath,amssymb,amsfonts}	
\usepackage{bm,braket,array}
\usepackage{multirow}
\usepackage[all]{xy}
\usepackage{amsthm}
\usepackage{amscd}
\usepackage{slashed} 
\usepackage{calligra} 
\usepackage[normalem]{ulem} 

\theoremstyle{definition}

\theoremstyle{remark}

\def\mod{{\rm\ mod\ }}

\def\tr{{\rm tr\,}}

\def\ua{\uparrow}
\def\da{\downarrow}

\newcommand{\s}{\sigma}

\newcommand{\la}{\lambda}

\newcommand{\C}{\mathbb{C}}

\newcommand{\R}{\mathbb{R}}
\newcommand{\Z}{\mathbb{Z}}

\def\widebar{\accentset{{\cc@style\underline{\mskip10mu}}}} 
\def\wideubar{\underaccent{{\cc@style\underline{\mskip10mu}}}} 

\begin{document}
\title{Higher Berry curvature from matrix product states}
\author{Ken Shiozaki}
\affiliation{Center for Gravitational Physics and Quantum Information, Yukawa Institute for Theoretical Physics, Kyoto University, Kyoto 606-8502, Japan}
\author{Niclas Heinsdorf}
    \affiliation{Max Planck Institute for Solid State Research, Heisenbergstrasse 1, 70569 Stuttgart, Germany}
    \affiliation{Stewart Blusson Quantum Matter Institute, University of British Columbia, Vancouver, BC V6T 1Z4, Canada}
    \affiliation{Department of Physics \& Astronomy, University of British Columbia, Vancouver, BC V6T 1Z1, Canada}  
\author{Shuhei Ohyama}
\affiliation{Center for Gravitational Physics and Quantum Information, Yukawa Institute for Theoretical Physics, Kyoto University, Kyoto 606-8502, Japan}
\date{\today}
\preprint{YITP-23-52}

\begin{abstract}
The higher Berry curvature was introduced by Kapustin and Spodyneiko as an extension of the Berry curvature in quantum mechanical systems with finite degrees of freedom to quantum many-body systems in finite spatial dimensions. 
In this paper, we propose an alternative formulation of the higher Berry curvature using translationally invariant matrix product states. 
They are the ground states of a set of gapped Hamiltonians which are evolved adiabatically through a discretized parameter space. 
Because matrix product states transform under a projective representation, evaluating the Berry curvature on a closed loop through parameter space is not sufficient to fix all the gauge degrees of freedom. 
To obtain a gauge-invariant real quantity, the higher-dimensional Berry curvature is evaluated on small tetrahedra in parameter space. 
Our numerical calculations confirm that the higher Berry curvature varies continuously throughout an adiabatic evolution and becomes quantized over a closed 3-dimensional parameter space.

\end{abstract}
\maketitle
\parskip=\baselineskip

\section{Introduction}

In a quantum mechanical system with $N$-degrees of freedom, a normalized pure state $\ket{\psi} \in \C^N, \braket{\psi|\psi}=1$, has an inherent $U(1)$ phase ambiguity. The physical state is given as the equivalence class that identifies this $U(1)$ phase ambiguity, which is often referred to as gauge ambiguity. For a single state $\ket{\psi}$, the gauge ambiguity is not a real degree of freedom because physically meaningful quantities are gauge-invariant. However, for a family of non-degenerate quantum states on a closed loop through parameter space, a unique $U(1)$ value can be defined. This quantity is called geometric or Berry phase~\cite{Berry}. It is a purely geometrical effect that arises from the non-commutativity of quantum operators and has been used to explain a variety of physical phenomena such as topological Hall effects, orbital magnetization or polarization~\cite{xiao2010berry,resta2007theory}.

Consider a one-parameter family of quantum states $\ket{\psi(x \in C)}$ along a closed loop $C = \{x(t) \in M | t \in [0,1], x(1)=x(0)\}$ in a parameter space $M$. Furthermore, we assume that the absolute value of the inner product $\braket{\psi(x(t))|\psi(x(t+\delta t))}$ is close to 1 between any two nearby points on the closed loop. Dividing the closed loop $C$ into $n$ intervals with $x_j=x(t=\frac{j}{n}), j=0,\dots,n$, 
the Berry phase $e^{i\gamma(C)}$ on that loop $C$ is given by
\begin{align}
\gamma(C)
&=
\lim_{n \to\infty} {\rm Arg} \prod_{j=0}^{n-1} \braket{\psi(x_{j})|\psi(x_{j+1})},
\label{eq:0D_BC}
\end{align}
where for a complex number $z=|z| e^{i\theta}$, ${\rm Arg}(z) = \theta, -\pi \leq \theta < \pi$.

The discrete expression (\ref{eq:0D_BC}) is manifestly gauge invariant. 
In practice, it is sufficient to take a large number of divisions $n$ so that $\gamma$ converges. 
The non-zero inner product $\braket{\psi(x(t))|\psi(x(t+\delta t))}$ assumes the absence of a transtion of symmetry charge along the closed loop $C$: 
If the state $\ket{\psi(x)}$ on the closed loop $C$ has some symmetry, $\ket{\psi(x)}$ is a one-dimensional representation of the symmetry group $G$ at each point. 
The absence of the transition assumes that the one-dimensional representation is constant on the closed loop.\footnote{
Consider a linear unitary representation $\hat g$ in the Hilbert space $\C^N$ for some symmetry group $G$. When the state $\ket{\psi(x \in M)}$ has $G$ symmetry at each point $x \in M$, $e^{i\theta_g(x)}=\braket{\psi(x)|\hat g|\psi(x)}, g \in G$ is a linear one-dimensional representation of $G$. If there exists some $h \in G$ such that $e^{i\theta_h(x(t))} \neq e^{i\theta_h(x(t+\delta t))}$, then $\braket{\psi(x(t))|\psi(x(t+\delta t))}=0$, and the phase cannot be defined. See Appendix\ref{app:Gsym_transfer_mat} for the extension to MPS.
}
Furthermore, the existence of internal degrees of freedom is essential for the Berry phase: 
For $N=1$, we always have $e^{i\gamma(C)}=1$. 
The minimum dimension giving a nontrivial Berry phase is $N=2$, and concretely, for $\ket{\psi(t)}=(\cos \frac{\theta}{2},\sin \frac{\theta}{2}e^{2\pi i t})^T,t \in [0,1]$, we have $\gamma(C) = \pi (1-\cos \theta) \mod 2\pi$.

The Berry phase for the boundary of a triangle $\Delta^2 = (x_0x_1x_2)$ in the parameter space,
\begin{align}
F(\Delta^2) &= {\rm Arg} \big[ \braket{\psi(x_0)|\psi(x_1)} \nonumber\\
&\quad \braket{\psi(x_1)|\psi(x_2)} \braket{\psi(x_2)|\psi(x_0)} \big],
\end{align}
corresponds to the magnetic flux piercing the triangle $\Delta^2$.
Note that the Berry phase $F(\Delta^2)$ is originally defined as a $\R/2\pi \Z$-value, but if the triangle $\Delta^2$ is so small that $e^{i F(\Delta^2)}$ is close to 1, $F(\Delta^2)$ can be considered as an $\R$-valued quantity. 
Consider a closed and oriented surface $\Sigma \subset M$ in which the modulus of the inner product $\braket{\psi(x)|\psi(x+\delta x)}$ between any two nearby points is close to 1.
Fix a triangulation $|\Sigma|$ of $\Sigma$.
The total sum of the magnetic flux $F(\Delta^2)$ on $\Sigma$ is manifestly quantized for any triangulation~\cite{Fukui_Hatsugai_Suzuki}:
\begin{align}
\nu(|\Sigma|) = \frac{1}{2\pi} \sum_{\Delta^2 \in |\Sigma|} F(\Delta^2) \in \Z.
\end{align}
If the triangulation $|\Sigma|$ is sufficiently fine with respect to the changes of the state $\ket{\psi(x)}$ on $\Sigma$, the integer value $\nu(|\Sigma|)$ is independent of the triangulation and called the Chern number $ch_1$. 
The Chern number $ch_1$ signifies an obstruction that prevents the state $\ket{\psi(x \in \Sigma)}$ from being represented by a single gauge on the closed surface $\Sigma$.

In this paper, we discuss an extension of the established Berry phase $\gamma(C)$ and the invariant $\nu(|\Sigma|)$ for finite-degree quantum mechanical systems to one-dimensional infinite many-body systems.
A naive extension would be to impose the periodic boundary condition with a system size $L$ and consider the Berry phase of the finite system as described above. 
However, in a one-dimensional system, the inner product between physically different states decays exponentially for $L$, so it is difficult to formulate the Berry phase in this approach. 
Instead, Kapustin and Spodyneiko have proposed the higher Berry curvatures that take values in $(d+2)$-forms for a family of unique gapped ground states in $d$-dimensional systems as a correlation function of the local Hamiltonian and its external differential in the parameter space~\cite{KapustinSpodyneiko_Berry_curvature20}.
See also \cite{Kapustin_Sopenko_Local_Noether_22}. 
In particular, for one-dimensional systems, the higher Berry curvature is a 3-form. 
The higher Berry curvature proposed by Kapustin and Spodyneiko is defined for any unique gapped ground state. 
Still, the calculated results are only available for the free fermionic model~\cite{KapustinSpodyneiko_Berry_curvature20} and spin systems that reduce to the two-site problem~\cite{Xueda}.

In this study, as one of the attempts to formulate the higher Berry curvature in one-dimensional systems with translational symmetry, we discuss an approach using the matrix product state (MPS) representation~\cite{MPS}.
In particular, towards applications in numerical calculations, we develop a formulation for higher Berry curvature in cases where the bond dimension of MPS varies in the parameter space. 
We will see that weighting by Schmidt eigenvalues plays an important role.
We also present numerical results obtained using infinite density matrix renormalization group (iDMRG).

During the preparation of this paper, \cite{Artymowicz_Kapustin_Sopenko_Quantization_23} appeared, where based on the formulation of \cite{Kapustin_Sopenko_Local_Noether_22}, it was proved that higher Berry curvatures in one-dimensional systems have integral periods.

The outline of the paper is as follows: 
Sec.~\ref{sec:Some background} discusses the background and motivates using MPS to construct higher Berry curvature.
Sec.\ref{sec:Gauge structure of MPS} is about the gauge structure of MPS.
Readers interested only in the formulation of the higher Berry curvature using MPS may skip Sec.\ref{sec:Some background} and Sec.~\ref{sec:Gauge structure of MPS}.
In Sec.~\ref{sec:Higher Berry curvature} we define the higher Berry curvature using the eigenvectors of the mixed transfer matrix from the MPS at each vertex of the discretized parameter space.
Sec.~\ref{sec:Example of Numerical Calculation} presents numerical results for the higher Berry curvature using iDMRG on a model based on \cite{Xueda} with added perturbations.
Sec.~\ref{sec:conc} provides the conclusion.
Several technical details are presented in two appendices.

\section{Some background}
\label{sec:Some background}

In this section, we provide the reader with some background that motivated this study. 

A conjecture on the topological structure of the space of invertible states has been proposed by Kitaev~\cite{Kit11,Kit13,Kit15}. 
Here, an invertible state refers to a pure state in spatial dimension $d$ that is realized as a gapped ground state of some local Hamiltonian, and for any sufficiently large set of site numbers $(L_{x_1},\dots,L_{x_d})$, there is no degeneracy in the ground state.

Let $F_d$ denote the space of all physical invertible states in $d$-spatial dimension. 
Here, ``physical" refers to the space obtained by removing any non-physical degrees of freedom, such as the $U(1)$ phase ambiguity of quantum states. 
It is conjectured that the homotopy equivalence exists between the space $F_d$ of invertible states in $d$-dimensions and the based loop space $\Omega F_{d+1}$ of invertible states in $(d+1)$-dimensions. 
\begin{align}
F_d 
&\cong \Omega F_{d+1} \nonumber\\
&= \{\ell: S^1 \to F_{d+1} | \ell(0)=\ell(1) = *\}. \label{eq:homoeq}
\end{align}
Here, the base point $*$ denotes the trivial tensor product state. 
In other words, this conjecture states that the sequence of spaces $\{F_d\}_{d \in \Z}$ forms an $\Omega$-spectrum in the generalized cohomology theory. 
The loop space can be physically interpreted as the space of adiabatic cycles, with the initial and final points being the trivial tensor product state.
As the generalization of the Thouless charge pump~\cite{Tho83} for any dimension and any on-site symmetry, it is known that in $(d+1)$-dimensional system, adiabatic cycles pump a $d$-dimensional invertible state at the boundary for each cycle, supporting the homotopy equivalence (\ref{eq:homoeq}) above.
See also \cite{TeoKane, GT17, Xio18, KSXiongGomi18, HsinKapustinThorngren, KS_adiabatic, bachmann2022classification, YuanOshikawaFurusaki_Gappability22, AasenWangHastings22, Xueda, Hsin_Wang23, ohyama2023discrete, beaudry2023homotopical} for related topics.

Let us consider the homotopy equivalence (\ref{eq:homoeq}) for spin systems without symmetry. 
First, in a 0-dimensional system, i.e., a quantum mechanical system with a finite number of degrees of freedom $N$, $F_0$ is nothing but a collection of lines in the complex vector space $\C^N$. 
Therefore, $F_0$ can be regarded as the complex projective space $\C P^{N-1} = \{\ket{\psi} \in \C^N | \braket{\psi|\psi}=1 \}/ (\ket{\psi} \sim e^{i\alpha} \ket{\psi})$.
Furthermore, by imposing stability for adding degrees of freedom to the space $F_0$, the wave function of a system with $N$ degrees of freedom can be embedded in a larger system with $(N+1)$ degrees of freedom as $\ket{\psi} = (\psi_1,\dots,\psi_N)^T \to (\psi_1,\dots,\psi_N,0)^T \in \C^{N+1}$. 
Taking the limit of embedding, we obtain an infinite-dimensional complex projective space $\C P^\infty = BU(1) \cong K(\Z,2)$ for $F_0$. 
Here, $BU(1)$ is the classifying space of the $U(1)$ group, where $U(1)$ corresponds to the phase ambiguity of the state. 
For a commutative group $G$, $K(G,n)$ is the Eilenberg-Maclane space characterized by $\pi_n(K(G,n))=G$ and $\pi_{m \neq n}(K(G,n))=0$~\cite{hatcher2002algebraic}.

$F_0 \sim K(\Z,2)$ and a candidate for $F_1$ that does not contradict relation (\ref{eq:homoeq}) is the Eilenberg-MacLane space $K(\Z,3)$, because $K(\Z,2) \cong \Omega K(\Z,3)$. 
Therefore, it is expected that the classification of 1-dimensional invertible states on the parameter space $M$ is given by the third cohomology $H^3(M,\Z) \cong [M,K(\Z,3)]$.
The space $K(\Z,3)$ is realized as the classifying space $BPU(\infty)$ of the projective unitary group $PU({\cal H})= U({\cal H})/(u \sim e^{i\alpha}u)$ of the infinite-dimensional Hilbert space ${\cal H}$.
As we will see in Sec.~\ref{sec:Gauge structure of MPS}, the gauge group of the MPS with fixed bond dimension $D$ is given by the projective unitary group $PU(D)$. 
The space $K(\Z,3)$ can be regarded as the classifying space of ``MPS with infinite bond dimension."

On the other hand, it is known that the classification of $PU(D)$ bundles is given by the $D$-torsion subgroup $\{x \in H^3(M,\Z)| Dx = 0\}$ of the cohomology group $H^3(M,\Z)$~\cite{Grothendieck1964-1966}. 
Therefore, an element of the integer group $\Z$ in $H^3(M,\Z)$ cannot be realized by a family of MPS with a constant bond dimension in the parameter space. 
However, a constant bond dimension $D$ in the whole parameter space and all $D$ Schmidt eigenstates contributing to physical quantities equally is quite a special situation.
In MPS, the bond dimension is a function that depends on the error between the exact quantum state and the number of sites, and the truncation of unessential Schmidt eigenvalues must not affect physical quantities. 
A generic family of MPS on a parameter space can be regarded as having an infinite bond dimension weighted by Schmidt eigenvalues, suggesting the realization of a non-trivial element in the integer group $\Z$ of $H^3(M,\Z)$ by a family of MPS. 
This is consistent with the model discussed in \cite{Xueda} with a non-trivial $\Z$ invariant by Kapustin and Spodyneiko~\cite{KapustinSpodyneiko_Berry_curvature20}, where the model is designed to change the bond dimension from $1$ to $2$ in the parameter space. 
See App.\ref{app:MPS_Xueda} for an MPS representation of the model discussed in \cite{Xueda}.

\section{Gauge structure of MPS}
\label{sec:Gauge structure of MPS}
We review the basic aspects of gauge degrees of freedom of MPS, and then investigate the properties of the transition function characterizing families of MPS on the parameter space. 
This section assumes a strictly constant bond dimension on the parameter space. 
In Sec.~\ref{sec:Higher Berry curvature}, we will extend the transition function of MPS bundles to the situation where the bond dimension changes.

\subsection{Injective and canonical MPS}
We refer to a set of $D \times D$ matrices $\{A^i\}_{i=1}^d$, $A^i \in {\rm Mat}_{D \times D}(\C)$, simply as an MPS, without distinguishing it from the corresponding quantum state (defined in (\ref{eq:mps_Lsites}), below). 
When dealing with different matrix sizes $D$ simultaneously, to specify the matrix size $D$, write $D$-MPS.
The bond dimension is denoted by $D$. 
The integer $d$ is the dimension of the local Hilbert space.

An MPS is called canonical~\cite{MPS} if it satisfies the right-canonical condition 
\begin{align}
\sum_{i=1}^d A^i A^{i\dag} = 1_D
\label{eq:right_cano}
\end{align}
and there exists a positive definite diagonal matrix $\Lambda^2$,
\begin{align}
\tr[\Lambda^2]=1, \label{eq:normalization_Lambda}
\end{align}
such that the left-canonical condition
\begin{align}
\sum_{i=1}^d A^{i\dag} \Lambda^2 A^i = \Lambda^2
\label{eq:left_cano}
\end{align}
is satisfied. 
Here, the entries of the matrix $\Lambda$, $\Lambda = {\rm diag}(\la_1,\la_2,\dots,)$, are the Schmidt eigenvalues, and the entanglement entropy for the bipartition of the left and right regions (in an open chain) are given by $S_E = -\sum_{j=1}^D \la_j^2 \log \la_j^2$.

The transfer matrix $T \in {\rm End}({\rm Mat}_{D \times D}(\C))$ is defined as 
\begin{align}
T(X) := \sum_{i=1}^d A^i X A^{i\dag}. 
\end{align}
We denote the spectral radius of $T$ by ${\rm spr}(T)$. 
An MPS is called injective~\cite{MPS} if the largest eigenvalue $\eta_0$ of the transfer matrix $T$ in the modulus, namely $|\eta_0| = {\rm spr}(T)$, is unique, the dimension of the corresponding eigenspace is 1, and the eigenvector $X_0$, $T(X_0) = \eta_0 X_0$, can be positive definite, i.e.,  $X_0>0$. 
In this case, the correlation length is given by $\xi=-1/\log(|\eta_1|/|\eta_0|)$ with $\eta_1$ the second largest eigenvalue of $T$ in modulus.~\footnote{
When the long-distance behavior of the correlation length is $\braket{{\cal O}_j {\cal O}'_{j'}} - \braket{{\cal O}_j}\braket{{\cal O}'_{j'}} \sim e^{-|j-j'|/\xi}$ for two local operators ${\cal O}_j, {\cal O}'_{j'}$, $\xi$ is called the correlation length.}
The parameter region where $|\eta_1|=|\eta_0|$ is satisfied corresponds to the phase transition point. 
The following discussion assumes the absence of the phase transition: 
There exists a constant finite gap $|\delta \eta|>0$ over the parameter space such that $|\eta_1|<|\eta_0|-|\delta \eta|$ holds. 
The transfer matrix $T$ is a completely positive map, so $\eta_0$ is known to be a positive real number. Moreover, for the injective MPS $\{A^i\}_{i=1}^d$, one can always satisfy the right-canonical condition (\ref{eq:right_cano}) by replacing
\begin{align}
A^i \to \frac{1}{\sqrt{\eta_0}} X_0^{-1/2} A^i X_0^{1/2}.
\end{align}
Further, in this case, if the left eigenstate of the transfer matrix $T$ with eigenvalue $\eta_0=1$ is denoted by $Y_0$, i.e., $\sum_{i=1}^d A^{i\dag} Y_0 A^i = Y_0$, then $Y_0$ is positive definite, and if the unitary matrix that diagonalizes $Y_0$ is $U \in U(D), Y_0=U^\dag \Lambda^2 U$, then by replacing
\begin{align}
A^i \to U A^i U^\dag, 
\end{align}
we can obtain the left-canonical condition (\ref{eq:left_cano}) while maintaining the right-canonical condition (\ref{eq:right_cano}).
Here we have used the spectral radius of the Hermitian conjugate of the transfer matrix $T^\dag(X) = \sum_{i=1}^d A^{i\dag}XA^i$ coincides with that of $T(X)$.

It is known that translation invariant states $\ket{\psi}$ satisfying periodic boundary conditions can be represented by a translation invariant MPS
\begin{align}
\ket{\{A^i\}_i}_L := \sum_{i_1,\dots,i_L}
\tr[A^{i_1}\cdots A^{i_L}] \ket{i_1\cdots i_L}
\label{eq:mps_Lsites}
\end{align}
The following theorem is sometimes called the fundamental theorem of the bosonic MPS~\cite{MPS}. 
Let $\{A_0^i\}_i$ and $\{A_1^i\}_i$ be injective $D$-MPSs in the canonical form, where the positive definite diagonal matrices in the left-canonical condition (\ref{eq:left_cano}) are given by $\Lambda_0^2$ and $\Lambda_1^2$, respectively. 
Suppose that there exists a site length $L > 2\ell + D^4$ and a $U(1)$ phase $e^{i \phi} \in U(1)$ such that $\ket{\{A_0^i\}_i}_L = e^{i\phi} \ket{\{A_1^i\}_i}_L$, i.e., the two MPSs represent the same physical state. 
(Here, $\ell$ is an integer determined by the MPS $\{A^i\}_i$.~\footnote{
When MPS $\{A^i\}_i$ is injective there exists an integer $\ell$ such that the set of products of matrices $\{A^{i_1}\cdots A^{i_\ell}\}_{i_1,\dots,i_{\ell}}$ span ${\rm Mat}_{D \times D}(\C)$ as a vector space~\cite{Sanz_WielandtsInequality}.
})
Then, there exist $e^{i\theta_{01}} \in U(1)$ and $V_{01} \in U(D)$ such that
\begin{align}
&A^i_0 = e^{i\theta_{01}} V_{01} A^i_{1} V_{01}^\dag,\label{eq:MPS_tr}\\
&V_{01} \Lambda_1 = \Lambda_0 V_{01},\label{eq:V_and_Lambda}
\end{align}
hold. 
Here, the phase factor $e^{i\theta_{01}}$ is unique, and $V_{01}$ is unique up to a $U(1)$ phase. In other words, the matrix $V_{01}$ is unique as an element of the projective unitary group $PU(D) = U(D)/U(1)$. 
As we will see in the next section, the pair of transition functions $(e^{i\theta_{01}}, V_{01})$ satisfies the cocycle condition for each, implying that a family of $D$-MPS on the parameter space $M$ defines a $U(1) \times PU(D)$ bundle~\cite{Haegeman_PUD_bundle14}.

Note that from the constraint (\ref{eq:V_and_Lambda}), the off-diagonal blocks of the transition function $V_{01}$ are zero between different Schmidt eigenvalues: 
The positive definite diagonal matrix $\Lambda$ can be chosen by a unitary transformation to satisfy $\la_1\geq \cdots \geq \la_D>0$ for the diagonal entries, while keeping the right-canonical condition (\ref{eq:right_cano}). 
In doing so for both $\Lambda_0$ and $\Lambda_1$, $\Lambda_0=\Lambda_1$ from (\ref{eq:V_and_Lambda}). 
By denoting the degeneracy of the Schmidt eigenvalue $\la_j$ as $D_j$ with $\sum_{j=1}^n D_j = D$, the transition function $V_{01}$ is block-diagonalized to a $U(D_1) \times \cdots \times U(D_n)$ matrix. 
In particular, the following holds: 
If there are no crossings between different blocks of Schmidt eigenvalues throughout the parameter space $M$, then the family of MPS over $M$ is a $P[U(D_1) \times \cdots \times U(D_n)] = [U(D_1)\times \cdots U(D_n)]/U(1)$ bundle.
This also implies that a nontrivial element of the free part of $H^3(M,\Z)$ is never obtained unless a crossing of Schmidt eigenvalues occurs.
See Sec.~\ref{sec:Cocycle condition}.  

From now on, unless otherwise specified, MPSs are assumed to be injective and satisfy the canonical conditions (\ref{eq:right_cano}) and (\ref{eq:left_cano}).

The transition function $V_{01}$ is computed as an eigenvector of the mixed transfer matrix introduced below~\cite{Perez-Garcia_Wolf_Sanz_Verstraete_Cirac_String_Order08}.
For two gauge-equivalent $D$-MPSs $\{A^i_0\}_i$ and $\{A^i_1\}_i$, define the mixed transfer matrix $T_{01} \in {\rm End}({\rm Mat}_{D \times D}(\C))$ as
\begin{align}
T_{01}(X)
:= \sum_{i=1}^d A^i_0 X A^{i\dag}_1.
\label{eq:minxed_TM}
\end{align}
From (\ref{eq:MPS_gauge}), we have
\begin{align}
T_{01}(X V^\dag_{01})
=e^{-i\theta_{01}}\sum_{i=1}^d A^i_0 X V^\dag_{01}A^{i\dag}_1.
\end{align}
Thus, the sets of eigenvalues of the transfer matrix $T_{00}(X)=\sum_{i=1}^d A^i_0 X A^{i\dag}_0$ and the mixed transfer matrix $T_{01}$ are in one-to-one correspondence: 
If $X_n$ is the eigenvector of the transfer matrix $T_{00}$ with eigenvalue $\eta_n$, then the corresponding eigenvalue and eigenvector of $T_{01}$ are given by $e^{-i\theta_{01}} \eta_n$ and $X_n V_{01}$ respectively.
In particular, the transition function $V_{01}$ is given as the eigenvector of the mixed transfer matrix $T_{01}$ corresponding to the largest eigenvalue in modulus.

In Sec.~\ref{sec:Higher Berry curvature}, we introduce the mixed transfer matrix between physically different MPS with different bond dimensions. 
The mixed transfer matrix plays an important role in formulating the higher Berry curvature.

\subsection{Cocycle condition}
\label{sec:Cocycle condition}
To construct an invariant in $H^3(M,\Z)$ from the data of transition functions consider a family of $D$-MPS on the parameter space $M$. 
Let $\{U_\alpha\}_\alpha$ be a good covering of $M$. 
We denote the intersections as $U_{\alpha_1\cdots \alpha_n}=U_{\alpha_1}\cap \cdots \cap U_{\alpha_n}$. 
The family of $D$-MPS in each patch $x\in U_\alpha$ is denoted as $\{A_\alpha^i(x)\}_i$, and let the positive definite diagonal matrix in the left-canonical condition (\ref{eq:left_cano}) be $\Lambda_\alpha(x)$. 
The gauge transformations (\ref{eq:MPS_tr}) hold in the intersection $U_{\alpha\beta}$ as
\begin{align}
&A^i_\alpha(x) = e^{i\theta_{\alpha\beta}(x)} V_{\alpha\beta}(x) A^i_\beta(x) V^\dag_{\alpha\beta}(x),
\label{eq:MPS_gauge}\nonumber\\
&\Lambda_\alpha(x) V_{\alpha\beta}(x) = V_{\alpha\beta}(x) \Lambda_\beta(x),
\end{align}
from which the $U(1)$ phase $e^{i\theta_{\alpha\beta}(x)}$ and the unitary matrix $V_{\alpha\beta}(x)$ are determined on $U_{\alpha\beta}$. 
Note that the $U(1)$ phase of $V_{\alpha\beta}(x)$ is undetermined. 
Repeating the patch transformations twice, we have
\begin{align}
A^i_\alpha(x)
&=e^{i\theta_{\alpha\beta}(x)}e^{i\theta_{\beta\alpha}(x)}\nonumber\\
&V_{\alpha\beta}(x) V_{\beta\alpha}(x) A^i_\alpha(x) V^\dag_{\beta\alpha}(x) V^\dag_{\alpha\beta}(x)
\end{align}
in $x\in U_{\alpha\beta}$, so from the uniqueness of the largest eigenvalue and corresponding eigenvector,
\begin{align}
e^{i\theta_{\beta\alpha}(x)}=e^{-i\theta_{\alpha\beta}(x)},
\end{align}
and $V_{\alpha\beta}(x)V_{\beta\alpha}(x)$ is equal to the identity matrix $1_D$ up to a $U(1)$ phase. 
From now on, we fix the $U(1)$ phase as
\begin{align}
V_{\beta\alpha}(x) = V^\dag_{\alpha\beta}(x). 
\end{align}
In the three patch intersection $x\in U_{\alpha \beta \gamma}$, by repeatedly performing gauge transformations (\ref{eq:MPS_tr}),
\begin{align}
A^i_\alpha(x)
&=e^{i\theta_{\alpha\beta}(x)}e^{i\theta_{\beta\gamma}(x)}e^{i\theta_{\gamma\alpha}(x)}\nonumber\\
&\quad V_{\alpha\beta}(x)V_{\beta\gamma}(x)V_{\gamma\alpha}(x)
A^i_\alpha(x) \nonumber\\
&\quad V^\dag_{\gamma\alpha}(x) V^\dag_{\beta\gamma}(x)V^\dag_{\alpha\beta}(x)
\end{align}
is obtained. 
Again, from the uniqueness of the largest eigenvalue and corresponding eigenvector,
\begin{align}
e^{i\theta_{\alpha\beta}(x)}e^{i\theta_{\beta\gamma}(x)}=e^{i\theta_{\alpha\gamma}(x)},
\label{eq:MPS_u1_cocycle}
\end{align}
and there exists $e^{i\phi_{\alpha\beta\gamma}(x)}\in U(1)$ such that
\begin{align}
V_{\alpha\beta}(x) V_{\beta\gamma}(x) V_{\gamma\alpha}(x) =e^{i\phi_{\alpha\beta\gamma}(x)} \times 1_D
\label{eq:ephi_def}
\end{align}
holds. 
Eq.(\ref{eq:MPS_u1_cocycle}) means that $\{e^{i\theta_{\alpha\beta}(x)}\}_{\alpha\beta}$ is a \v{C}ech 1-cocycle that defines a $U(1)$ bundle, but it is a ``weak" object related to the zero-dimensional state originating from translational symmetry, and will not be discussed further in this paper.

Here, from the normalization condition (\ref{eq:normalization_Lambda}) on $\Lambda$s, the $U(1)$ phase $e^{i\phi_{\alpha\beta\gamma}(x)}$ is given by~\cite{OhyamaRyu2023higher}
\begin{align}
e^{i\phi_{\alpha\beta\gamma}(x)} 
= \tr[\Lambda^2_\alpha(x) V_{\alpha\beta}(x) V_{\beta\gamma}(x) V_{\gamma\alpha}(x)]
\label{eq:2cocyle_0}
\end{align}
Thus, only the dominant Schmidt eigenvalues contribute the $U(1)$ phase $e^{i\phi_{\alpha\beta\gamma}(x)}$. 
An equivalent expression that is symmetric is 
\begin{align}
e^{i\phi_{\alpha\beta\gamma}(x)} 
&= \tr\big[\Lambda^\frac{2}{3}_\alpha(x) V_{\alpha\beta}(x) \nonumber \\
&\quad \Lambda^\frac{2}{3}_\beta(x) V_{\beta\gamma}(x) \Lambda^\frac{2}{3}_\gamma(x) V_{\gamma\alpha}(x)\big].
\label{eq:2cocycle_cont}
\end{align}

For each patch $U_\alpha$, the gauge transformation
\begin{align}
&A_\alpha(x) \mapsto e^{i\chi_\alpha(x)} W_\alpha(x)A_\alpha(x)W^\dag_\alpha(x), \\
&\Lambda_\alpha(x) \mapsto W_\alpha(x)\Lambda_\alpha(x) W^\dag_\alpha(x), \label{eq:gauge_tr_Lambda}
\end{align}
with $x \in U_\alpha$, induces the following gauge transformation for the transition function $V_{\alpha\beta}(x)$. 
\begin{align}
V_{\alpha\beta}(x) \mapsto W_\alpha(x) V_{\alpha\beta}(x) W^\dag_\beta(x),
\quad x \in U_{\alpha\beta}.
\label{eq:gauge_tr_V}
\end{align}
Note that the $U(1)$ phase $e^{i\phi_{\alpha\beta\gamma}(x)}$ is gauge invariant under the gauge transformations (\ref{eq:gauge_tr_Lambda}) and (\ref{eq:gauge_tr_V}).

For the four patch intersection $U_{\alpha\beta\gamma\delta}$, the 2-cocycle condition
\begin{align}
e^{i \phi_{\beta\gamma\delta}(x)}e^{-i\phi_{\alpha\gamma\delta}(x)}e^{i\phi_{\alpha\beta\delta}(x)}e^{-i\phi_{\alpha\beta\gamma}(x)}=1
\label{eq:2cocyle}
\end{align}
follows from the two different decomposition of $V_{\alpha\beta}(x) V_{\beta\gamma}(x) V_{\gamma\delta}(x)$. Therefore, $\{e^{i\phi_{\alpha\beta\gamma}(x)}\}_{\alpha\beta\gamma}$ is a \v{C}ech 2-cocycle.
Moreover, the phase change of $V_{\alpha\beta}(x)$, $V_{\alpha\beta}(x) \mapsto V_{\alpha\beta}(x) e^{i\xi_{\alpha\beta}(x)}$, induces an equivalent relation
\begin{align}
e^{i\phi_{\alpha\beta\gamma}(x)} \sim
e^{i\phi_{\alpha\beta\gamma}(x)} \times
e^{i\xi_{\beta\gamma}(x)} e^{-i\xi_{\alpha\gamma}(x)} e^{i\xi_{\alpha\beta}(x)}.
\end{align}
Thus, the equivalence class $[e^{i\phi_{\alpha\beta\gamma}(x)}]$ determines an element of the \v{C}ech cohomology $\check H^2(M,\underline{U(1)})$, where $\underline{U(1)}$ is the sheaf of continuous $U(1)$-valued functions on the parameter space $M$.
Furthermore, fix a lift
\begin{align}
\R/2\pi \Z \ni \phi_{\alpha\beta\gamma}(x) \to \tilde \phi_{\alpha\beta\gamma}(x) \in \R
\label{eq:lift}
\end{align}
and take the differential 
\begin{align}
&c_{\alpha\beta\gamma\delta}
= \frac{1}{2\pi}(\delta \tilde \phi(x))_{\alpha\beta\gamma\delta} \nonumber\\
&= \frac{1}{2\pi}\left\{ \tilde \phi_{\beta\gamma\delta}(x) - \tilde \phi_{\alpha\gamma\delta}(x) + \tilde \phi_{\alpha\beta\delta}(x) - \tilde \phi_{\alpha\beta\gamma}(x)\right\}.\label{eq:3cocycle_def}
\end{align}
Since $\delta c=0$ and, due to the 2-cocycle condition (\ref{eq:2cocyle}), $c_{\alpha\beta\gamma\delta}$ is a constant integer over $U_{\alpha\beta\gamma\delta}$, $\{ c_{\alpha\beta\gamma\delta}\}_{\alpha\beta\gamma\delta}$ is a \v{C}ech 3-cocycle with integer coefficients. 
Also, the change of the lift (\ref{eq:lift}) induces an equivalence relation $c \sim c + \delta b$ with $b \in \check C^2(M,\Z)$, so eventually $c_{\alpha\beta\gamma\delta}$ determines an element of the third \v{C}ech cohomology group $\check H^3(M,\Z)$.
This is nothing but a explicit construction of the isomorphism $\check H^2(M,\underline{U(1)}) \cong \check H^3(M,\Z)$.

Note that when the bond dimension $D$ is constant over the whole parameter space $M$, it is shown that the element $[c]$ takes values in the $D$-torsion subgroup as follows~\cite{Grothendieck1964-1966}. 
Due to the $U(1)$ ambiguity, the transition function $V_{\alpha\beta}(x)$ can be chosen such that $\det V_{\alpha\beta}(x) = 1$, i.e., it takes values in the $SU(D)$ group. 
In this case, the ambiguity of the matrix $V_{\alpha\beta}(x)$ is discrete, given by the $D$-th roots of unity, $\Z_D = \{ e^{\frac{2\pi i p}{D}} \in U(1) | p=0,\dots, D-1\}$, and thus the 2-cocycle $e^{i\phi_{\alpha\beta\gamma}(x)}$ is independent of $x$ and determines an element of the cohomology group $\check H^2(M,\Z_D)$.
In this case, the 3-cocycle $c$ defined by (\ref{eq:3cocycle_def}) satisfies that $D c = \delta b$ for some $b$, and thus the 3-cocycle $c$ takes values in the $D$-torsion subgroup $\{[c] \in \check H^3(M,\Z) | D [c]=0\}$.

\section{Construction of higher Berry phase}
\label{sec:Higher Berry curvature}
In the previous section, we saw that a family of MPS with a constant bond dimension cannot realize the free part of the cohomology group $H^3(M,\Z)$. 
However, the bond dimension is a non-physical parameter as it is introduced for approximating quantum states, and in general the bond dimension varies throughout the parameter space $M$. 
This section considers the case where the bond dimension $D_x$ varies depending on the lattice point $x$ in a discretely approximated parameter space, with numerical computation of higher Berry curvature in mind.
First, we introduce the mixed transfer matrix between two physically different, i.e. gauge-inequivalent MPSs.

\subsection{Mixed transfer matrix}
Let $\{A^i_0\}_i$ and $\{A^i_1\}_i$ be a $D_0$-MPS and $D_1$-MPS, respectively, both injective and satisfying the canonical conditions (\ref{eq:right_cano}) and (\ref{eq:left_cano}).
Denote the positive definite diagonal matrices in the left-canonical form (\ref{eq:left_cano}) as $\Lambda_0$ and $\Lambda_1$.
We introduce the mixed transfer matrix $T_{01} \in {\rm End}({\rm Mat}_{D_0 \times D_1}(\C))$ in the same way as (\ref{eq:minxed_TM}):
\begin{align}
    T_{01}(X):=\sum_{i=1}^d A^i_0 X A^{i\dag}_1. 
    \label{eq:mixed_TM_01}
\end{align}
If the $D_0$-MPS $\{A^i_0\}_i$ and $D_1$-MPS $\{A^i_1\}_i$ are close to each other as physical states, the spectrum of the mixed transfer matrix $T_{01}$ is expected to show little variation compared to the transfer matrix between gauge-equivalent MPS.
In particular, (i) the largest eigenvalue $\eta_{01}$ of $T_{01}$ in modulus is unique, its eigenspace is one-dimensional, and (ii) there exists a finite $|\delta \eta|>0$ such that $|\eta|<|\eta_{01}|-|\delta \eta|$ holds for other eigenvalues $\eta$. 
We define that the MPS of $\{A_0^i\}_i$ and $\{A_1^i\}_i$ are close to each other as satisfying (i) and (ii).

If $\{A_0^i\}_i$ and $\{A_1^i\}_i$ belong to different symmetry-protected topological (SPT) phases, the eigenvalues of the mixed transfer matrix $T_{01}$ are necessarily degenerate in modulus. 
For more details, see Appendix~\ref{app:Gsym_transfer_mat}. 
The uniqueness of the largest eigenvalue of the mixed transfer matrix in modulus is a necessary condition for the two MPSs, $\{A_0^i\}_i$ and $\{A_1^i\}_i$, to belong to the same SPT phase for the symmetry group $G$.

\begin{figure}[!]
\centering
\includegraphics[width=0.8\linewidth, trim=0cm 0cm 0cm 0cm]{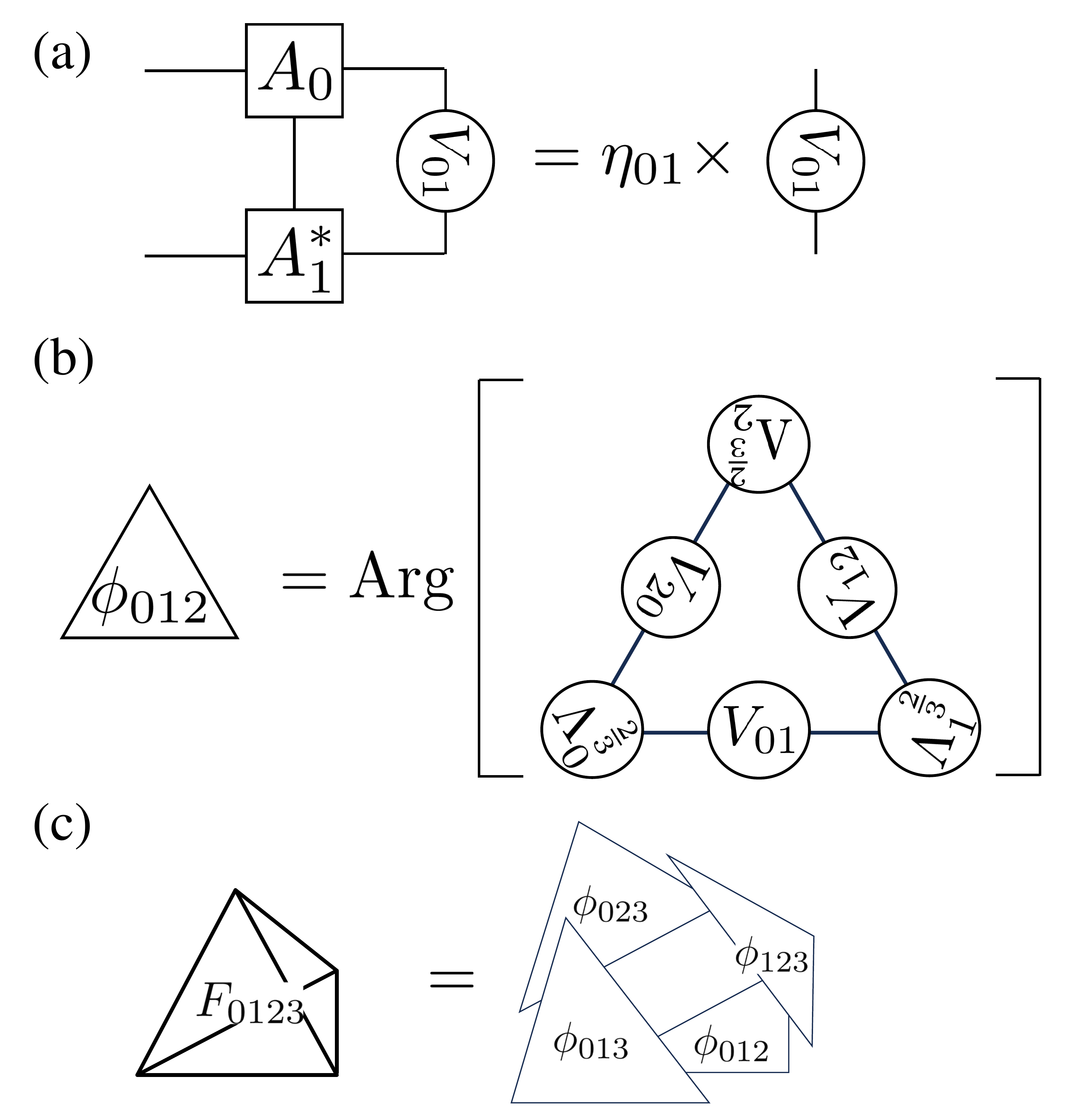}
\caption{
(a) The overlap matrix $V_{01}$ over a 1-simplex $\Delta^{(1)}=(01)$ is defined as an eigenvector of the mixed transfer matrix. 
(b) $\phi_{012}$ is the integrated higher Berry connection over a $2$-simplex $\Delta^{(2)}=(012)$. 
(c) The higher Berry flux over a 3-simplex $\Delta^{(3)}=(0123)$. 
}
\label{fig:tensor_diagram}
\end{figure}

Assuming that the two MPSs $\{A^i_0\}_i$ and $\{A^i_1\}_i$ are close, let the eigenvector corresponding to the largest eigenvalue of the mixed transfer matrix $T_{01}$ be $V_{01}$: 
\begin{align}
    &T_{01}(V_{01})=
    \sum_{i=1}^d A_0^i V_{01} A_1^{i\dag}
    =\eta_{01} V_{01}, \nonumber \\
    &|\eta_{01}|= {\rm spr}(T_{01}). \label{eq:V01_def}
\end{align}
See Fig.~\ref{fig:tensor_diagram} (a) for the diagrammatic expression. 
From the mixed transfer matrix (\ref{eq:mixed_TM_01}) between two given close MPSs, the $D_0 \times D_1$ matrix $V_{01}$ is uniquely determined up to a constant factor in this way. 
The Hermitian conjugate of (\ref{eq:V01_def}) is
\begin{align}
    \sum_{i=1}^d A_1^i V_{01}^\dag A_0^{i\dag} = \eta_{01}^* V_{01}^\dag, 
\end{align}
which implies that, up to a constant factor, $V_{10} \sim V_{01}^\dag$ holds.
From now on, we always assume that 
\begin{align}
    V_{10}=V_{01}^\dag. 
    \label{eq:V10toV01}
\end{align}

There are two types of gauge transformations in the matrix $V_{01}$ determined by (\ref{eq:V01_def}).
The first one is the gauge transformation of the MPS, $A^i_n \mapsto e^{i\theta_n} W_n A^i_n W^\dag_n, W_n \in U(D_n), n \in \{0,1\}$, which induces the transformation
\begin{align}
    \Lambda_n \mapsto W_n \Lambda_n W_n^\dag, \quad 
    V_{01} \mapsto W_0 V_{01} W^\dag_1. 
    \label{eq:1st_gauge}
\end{align}
The second one is the ambiguity of the constant factor of $V_{01}$ itself,
\begin{align}
    V_{01} \mapsto z_{01} V_{01},\quad z_{01} \in \C^{\times }. 
    \label{eq:2nd_gauge}
\end{align}
The higher Berry curvature must be constructed to be gauge-invariant under these two gauge transformations (\ref{eq:1st_gauge}) and (\ref{eq:2nd_gauge}).
Roughly speaking, the inner product $\braket{\psi_0|\psi_1}$ in the expression of the Berry phase (\ref{eq:0D_BC}) corresponds to the weighted transition function $\Lambda^{\alpha}_0 V_{01} \Lambda^{\alpha}_1$ with Schmidt eigenvalues $\Lambda$, where the weight $\alpha>0$ must be properly set.
Focusing on this similarity, we formulate the higher Berry phase in the following section.

\subsection{Higher Berry connection}
Consider a family of MPS on a parameter space $M$.
Take a discrete approximation $|M|$ of $M$ by triangulation and fix the orientations of the triangles. 
For any given triangle, two edges will have orientations opposite to the third edge.
We denote the set of vertices by $S_v$.
The bond dimension $D_p$ of the MPS can vary depending on the vertices $p \in S_v$.
At each vertex $p \in S_v$, we have an injective and canonical MPS $\{A^i_p\}_i$. 
Denote the positive definite diagonal matrices in the left-canonical condition (\ref{eq:left_cano}) as $\Lambda_p$.

For any edge $(p_0p_1)$, the $D_{p_0}$-MPS $\{A_{p_0}^i\}_i$ and the $D_{p_1}$-MPS $\{A_{p_1}^i\}_i$ are assumed to be uniformly close in the sense of the previous section.
That is, the largest eigenvalue $|\eta_{p_0p_1}|={\rm spr}(T_{p_0p_1})$ of the mixed transfer matrix $T_{p_0p_1}$ defined by (\ref{eq:mixed_TM_01}) is unique in modulus, and there exists a finite gap $|\delta \eta|$ independent of the vertex $p \in S_v$ such that for any other eigenvalue $\eta$ satisfies $|\eta|<|\eta_{p_0p_1}|-|\delta \eta|$ for any edges.
In particular, any two points on the parameter space $|M|$ are connected without phase transitions.

Introduce the following quantity, which takes values in $U(1)$, corresponding to the discrete approximation of (\ref{eq:2cocycle_cont}) for any triangle $\Delta^2 = (p_0p_1p_2)$ in the triangulation $|M|$:
\begin{align}
\phi(p_0p_1p_2)&:= {\rm Arg}\ \tr [\Lambda_{p_0}^\frac{2}{3} V_{p_0p_1} \Lambda_{p_1}^\frac{2}{3} V_{p_1p_2} \Lambda_{p_2}^\frac{2}{3} V_{p_2p_0}]\nonumber\\
&\in \R/2\pi \Z.
\label{eq:HBP_Delta^2}
\end{align}
See Fig.~\ref{fig:tensor_diagram} (b) for the diagrammatic expression. 
Clearly, $\phi(p_0p_1p_2)$ is invariant under the first gauge transformation (\ref{eq:1st_gauge}).

Invariance under the second gauge transformation (\ref{eq:2nd_gauge}) is obtained by summing up the phases $\phi(\Delta^2)$ for all triangles $\Delta^2 \in |\Sigma_2|$ on the two-dimensional closed surface $|\Sigma_2|$ in $|M|$.
Define the higher Berry phase on $|\Sigma_2|$ as
\begin{align}
    \gamma(|\Sigma_2|)
    := \sum_{\Delta^2 \in |\Sigma_2|} \sigma(\Delta^2) \phi(\Delta^2) \in \R/2\pi \Z, 
    \label{eq:HBP}
\end{align}
where the sign factor $\sigma(\Delta^2) \in \{\pm 1\}$ is defined as $\sigma(\Delta^2)=1$ when the orientation of $\Delta^2$ agrees with that of $\Sigma_2$, and $\sigma(\Delta^2)=-1$ when they do not agree.
With the constraint (\ref{eq:V10toV01}), the second gauge transformation (\ref{eq:2nd_gauge}) cancels the contributions from adjacent triangles, making $\gamma(|\Sigma_2|)$ a gauge-invariant quantity.
The quantity $\sigma(\Delta^2) \phi(\Delta^2)$ can be regarded as a discrete approximation of the integral of the higher Berry connection, which is a 2-form, over a small triangle $\Delta^2$.

Note that the expression
\begin{align}
    {\rm Arg}\ \tr [\Lambda_{p_0}^2 V_{p_0p_1} V_{p_1p_2} V_{p_2p_0}] \label{eq:bad1}
\end{align}
corresponding to the discrete approximation of (\ref{eq:2cocyle_0}) can also be considered as a candidate for the $U(1)$-valued $\phi(p_0p_1p_2)$. 
However, it turns out that the expression (\ref{eq:bad1}) is inappropriate, as the higher Berry curvature introduced in the next section depends on the bond dimension $D_p$.
Nevertheless, the sum of the higher Berry curvatures defined in (40) is always quantized.

\subsection{Higer Berry curvature}
For a tetrahedron $\Delta^3 = (p_0p_1p_2p_3)$ in $|M|$, the higher Berry phase (\ref{eq:HBP}) on its boundary $\partial \Delta^3 = (p_1p_2p_3) - (p_0p_2p_3) + (p_0p_1p_3) - (p_0p_1p_2)$ determines a gauge-invariant quantity. We consider this as the discrete approximation of the higher Berry curvature in the tetrahedron $\Delta^3$:
\begin{align}
F(\Delta^3)
&:=\gamma(\partial \Delta^3)\nonumber\\
&= \phi(p_1p_2p_3)-\phi(p_0p_2p_3) \nonumber\\
&+\phi(p_0p_1p_3)-\phi(p_0p_1p_2).
\label{eq:HBC}
\end{align}
See Fig.~\ref{fig:tensor_diagram} (c).
Although $F(\Delta^3)$ is originally valued in $\R/2\pi \Z$, when the four points $p_0p_1p_2p_3$ are sufficiently close, $F(\Delta^3)$ is close to 0 as an element of $\R/2\pi \Z$, so the $2\pi$ ambiguity does not exist, and $F(\Delta^3)$ can be considered as a real number. 
By design, the sum of the curvatures $F(\Delta^3)$ over an orientable 3-dimensional closed manifold $|\Sigma_3|$ in $|M|$ is quantized:
\begin{align}
\nu(|\Sigma_3|):= \frac{1}{2\pi} \sigma(\Delta^3) \sum_{\Delta^3 \in |\Sigma_3|} F(\Delta^3) \in \Z. 
\end{align}
Here, the sign $\sigma(\Delta^3) \in \{\pm 1\}$ takes the value $1$ if the orientation of the tetrahedron $\Delta^3$ coincides with that of $\Sigma_3$, and $-1$ otherwise. 
The integer value $\nu(|\Sigma_3|)$ can be thought of as characterizing the integer part of the 3rd cohomology $H^3(M,\Z)$. 
We calculate $\nu(|\Sigma_3|)$ for a specific example in the next section.

\section{A numerical calculation}
\label{sec:Example of Numerical Calculation}
In this section, we present a numerical calculation of the higher Berry curvature (\ref{eq:HBC}) for a model obtained by adding a perturbation to that analyzed in \cite{Xueda}.

\subsection{Model}

\begin{figure}[!]
\centering
\includegraphics[width=\linewidth, trim=2cm 0cm 2cm 0cm]{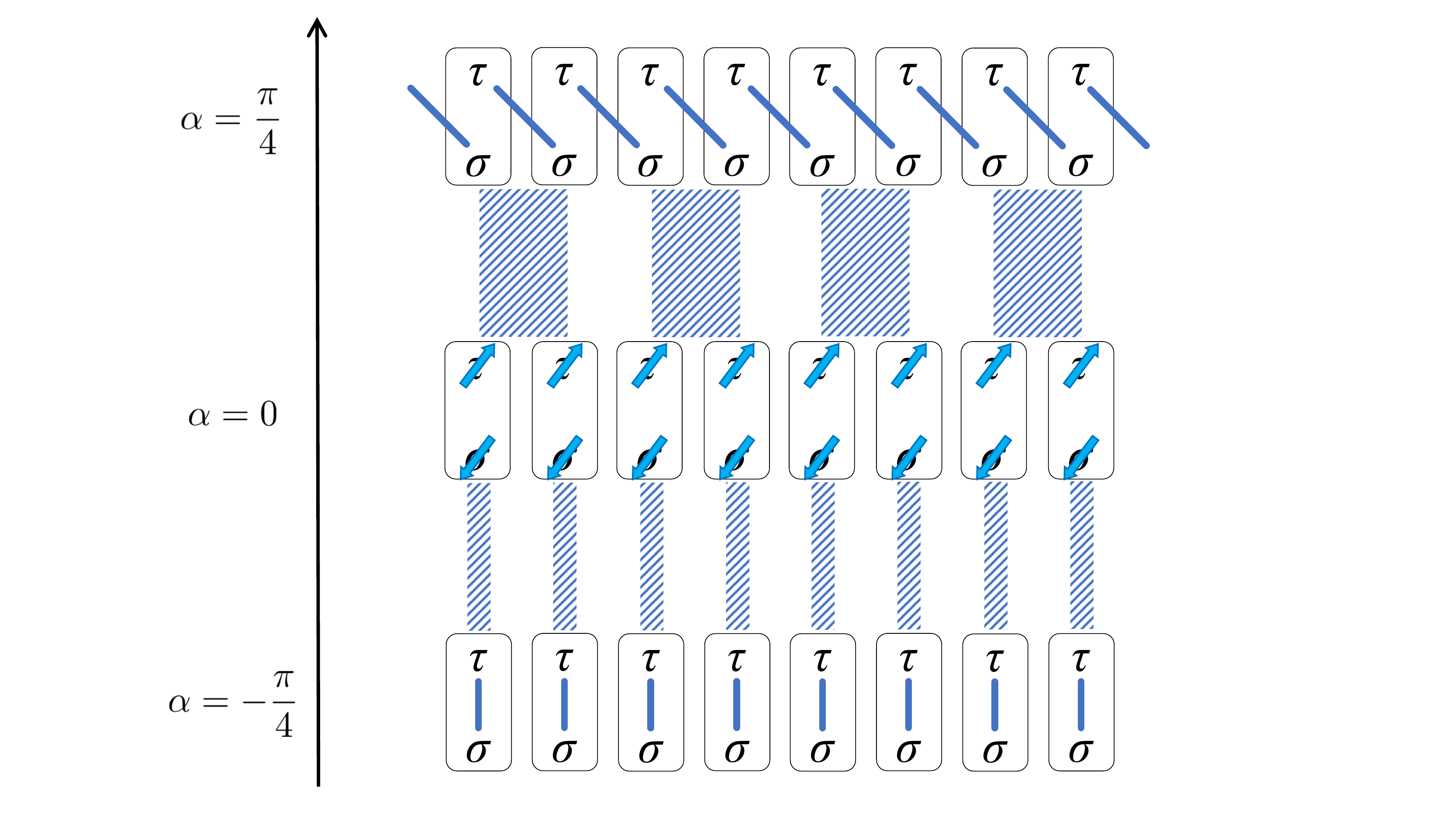}
\caption{The unperturbed Hamiltonian $H_0(\alpha,\bm{n})$.
The rods represent antiferromagnetic interactions, the arrows represent a magnetic field in the $\bm{n} \in S^2$ direction, and the hatching represents changes in the interaction in the $\alpha$ direction.
}
\label{fig:model}
\end{figure}

We take the parameter space to be $M=S^3$ and define the coordinates $\bm{w} = (w_0,w_1,w_2,w_3) \in S^3$ as
\begin{align}
    &w_0 = \cos (2\alpha),\quad (w_1,w_2,w_3) = \sin (2 \alpha) \bm{n}, \\
    &\alpha \in [-\frac{\pi}{4},\frac{\pi}{4}],\quad |\bm{n}|=1.
\end{align}
We introduce two spin-1/2 degrees of freedom at integer sites and denote the Pauli matrices for each as $\bm{\sigma}_l, \bm{\tau}_l$. 
First, we introduce the unperturbed Hamiltonian $H_0(\alpha,\bm{n})$~\cite{Xueda}:
\begin{align}
    H_0(\alpha,\bm{n}) 
    &= \sin (2\alpha) \sum_{l \in \Z} \left\{ \begin{array}{ll}
    -\bm{\sigma}_l \cdot \bm{\tau}_l & (\alpha \in [-\frac{\pi}{4},0]) \\
    \bm{\tau}_l \cdot \bm{\sigma}_{l+1} & (\alpha \in [0,\frac{\pi}{4}]) \\
    \end{array}\right. \nonumber \\
    &+ \cos (2\alpha) \sum_{l \in \Z} (-\bm{n} \cdot \bm{\s}_l + \bm{n} \cdot \bm{\tau}_l).
    \label{eq:H0}
\end{align}
The Hamiltonian $H_0(\alpha,\bm{n})$ is defined on the parameter space $S^3$ because the $\bm{n}$ dependence vanishes at $\alpha=-\pi/4, \pi/4$.
Although $H_0(\alpha,\bm{n})$ is not smooth at $\alpha=0$, it is continuous. 
This model is a 2-site problem and can be solved analytically; it has a finite energy gap between the ground state and the first excited state and no degeneracy in the ground state. 
In particular, for $\alpha \in[-\pi/4,0]$, the ground state is a tensor product state, and the bond dimension is 1.
For $\alpha \in [0,\pi/4]$, the adjacent sites form Bell pairs, and the bond dimension is 2.
In \cite{Xueda}, the higher Berry curvature by Kapustin and Spodyneiko~\cite{KapustinSpodyneiko_Berry_curvature20} is calculated, and it is confirmed that the integral over $S^3$ is quantized to 1.

The Hamiltonian $H_0(\alpha,\bm{n})$ is slightly modified from the expression in \cite{Xueda} to simplify the MPS notation. 
Although the analytical expression for MPS is not needed in the numerics, the MPS representation of the ground state of $H_0(\alpha,\bm{n})$ is provided in Appendix~\ref{app:MPS_Xueda}. 

As perturbations, we add the nearest-neighbor Heisenberg term and the next-nearest-neighbor Heisenberg term:
\begin{align}
    &V_1 = \sum_{l \in \Z} (\bm{\sigma}_l \cdot \bm{\tau}_l + \bm{\tau}_l \cdot \bm{\sigma}_{l+1}), \\
    &V_2 = \sum_{l \in \Z} (\bm{\sigma}_l \cdot \bm{\sigma}_{l+1} + \bm{\tau}_l \cdot \bm{\tau}_{l+1}). 
\end{align}
The model we compute the higher Berry curvature (\ref{eq:HBC}) is
\begin{align}
    H(\alpha,\bm{n})
    =H_0(\alpha,\bm{n})+J_1 V_1 + J_2 V_2. 
\end{align}
A peculiarity of the model $H(\alpha,\bm{n})$ is that a unitary transformation can obtain its $\bm{n}$ dependence. Let $\bm{n} = (\theta,\phi)$ be the polar coordinates, and introduce the unitary transformation $U(\bm{n})$ as
\begin{align}
    &U(\bm{n}) = \bigotimes_{l \in \Z} u_l(\bm{n}), \nonumber\\
    &u_l(\bm{n}) = e^{-i\phi \s^z_l/2} e^{-i \theta \s_l^y/2} e^{-i\phi \tau^z_l/2} e^{-i \theta \tau_l^y/2}.
    \label{eq:utr_for_S2}
\end{align}
Then, we have
\begin{align}
    H(\alpha,\bm{n})
    =U(\bm{n})H(\alpha,\bm{n}=\hat z)U(\bm{n})^{-1}
\end{align}
where $\hat z=(0,0,1)$. 
Thus, the $S^2$ dependence of the MPS can be obtained as
\begin{align}
    A^{\sigma\tau}(\alpha,\bm{n})=\sum_{\s',\tau'} [u(\bm{n})]_{\s'\tau'} A^{\s'\tau'}(\alpha,\hat z),
    \label{eq:so3_unitary_tr}
\end{align}
where $u(\bm{n})$ is the unitary matrix at a single site $u_l(\bm{n})$. 
Furthermore, the integrated higher Berry curvature (\ref{eq:HBC}) over $S^2$ can be calculated for a small area $\delta S$ of $S^2$ and multiplied by ${\rm Area}(S^2)/{\rm Area}(\delta S)$.

\subsection{Results}

In the following, we show the numerical calculations of the higher Berry curvature for two cases in the parameter space $(\alpha,\theta,\phi)$: one in which $(J_1, J_2)$ is fixed and the other in which $(J_1, J_2)$ is given randomly.

\subsubsection{Fixed $(J_1,J_2)$}

\begin{figure}[!]
\centering
\includegraphics[width=\linewidth, trim=1cm 2cm 1cm 1cm]{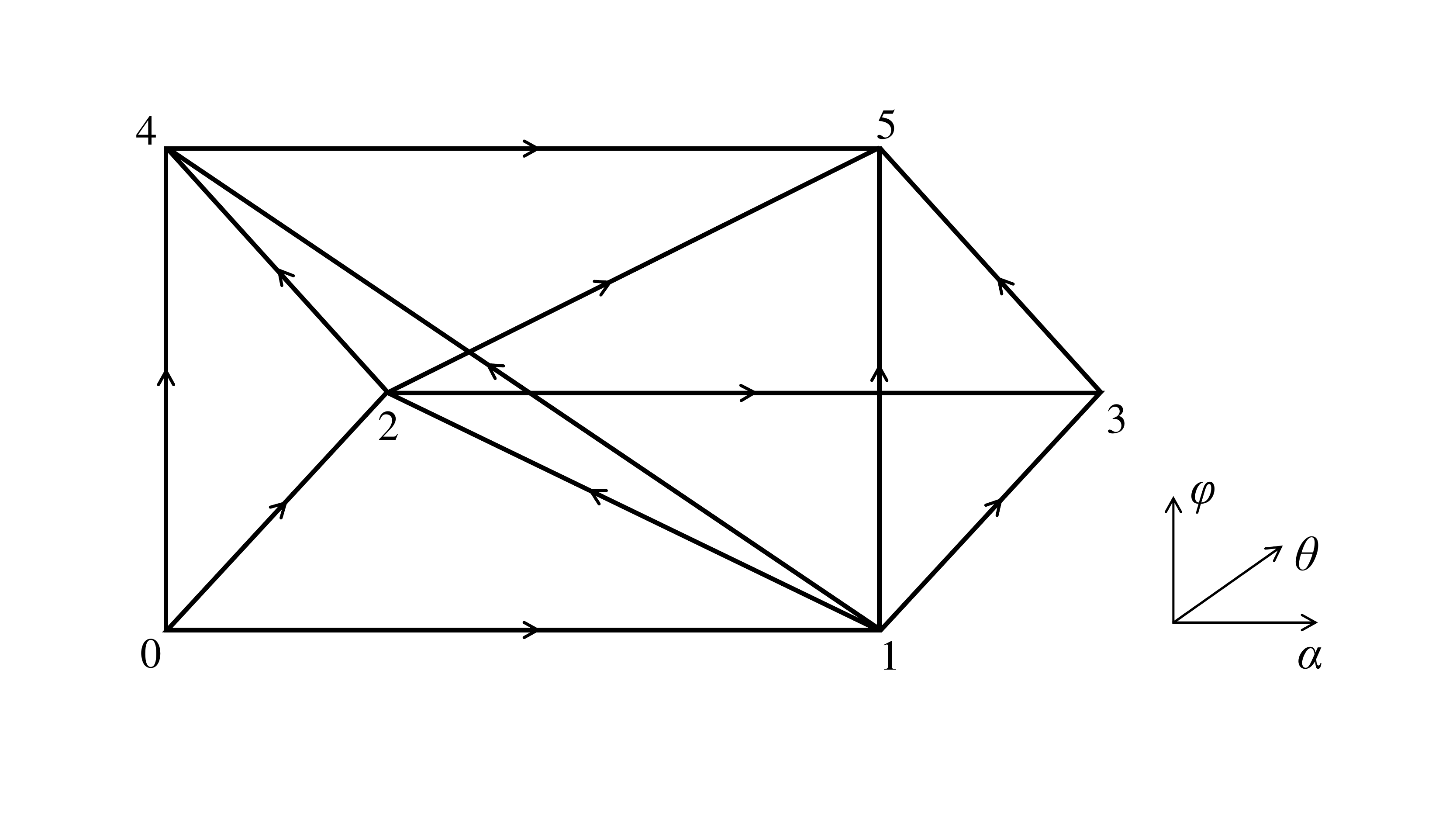}
\caption{
A small triangular prism $\Gamma$ and its triangulation in the spherical shell $[\alpha_j, \alpha_{j+1}]$. 
The orientation is given along the numbers.
}
\label{fig:cell}
\end{figure}

\begin{figure*}[!]
\centering
\begin{align*}
&\includegraphics[width=0.5\linewidth, trim=0cm 0cm 0cm 0cm]{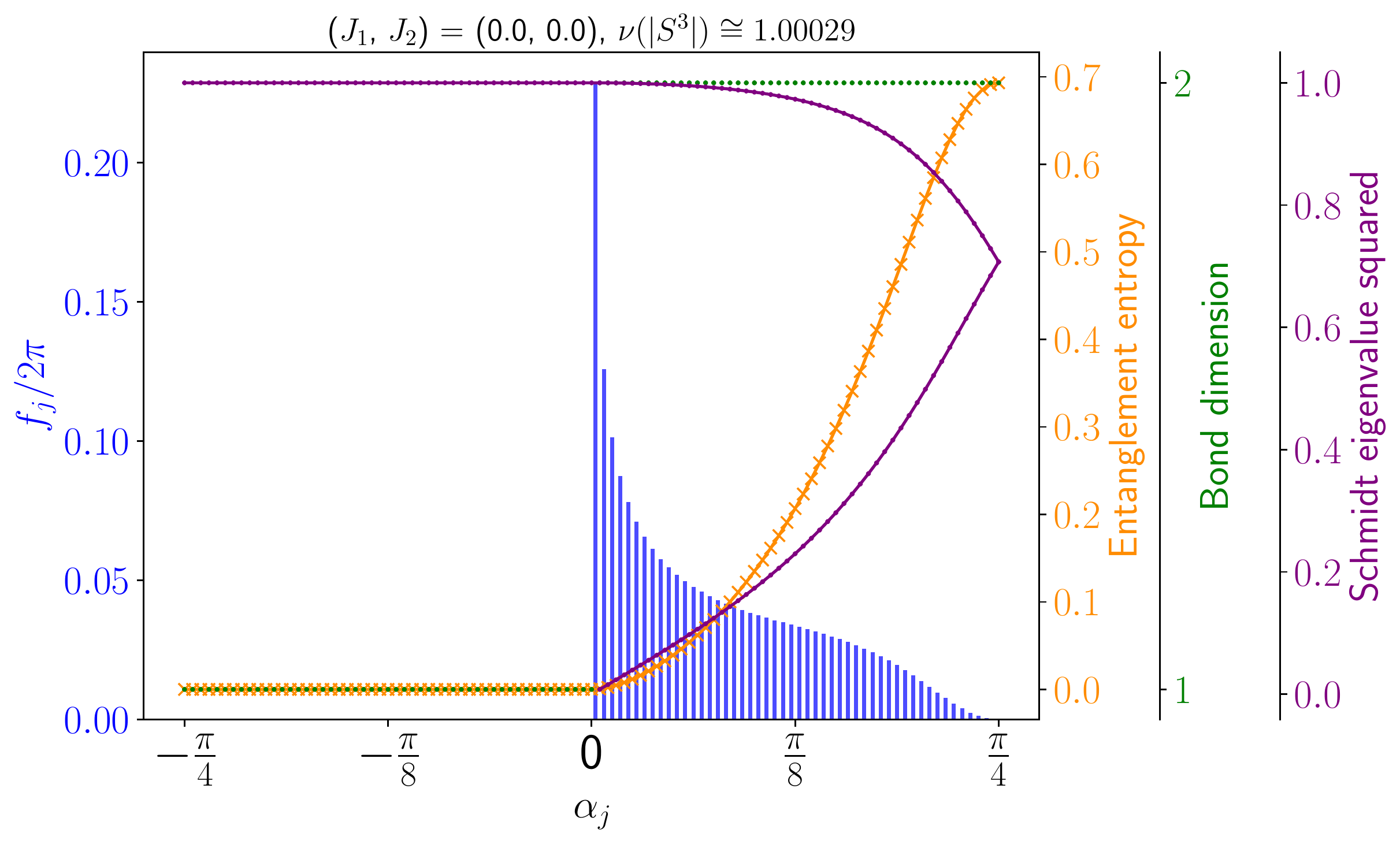} \quad & \quad 
\includegraphics[width=0.5\linewidth, trim=0cm 0cm 0cm 0cm]{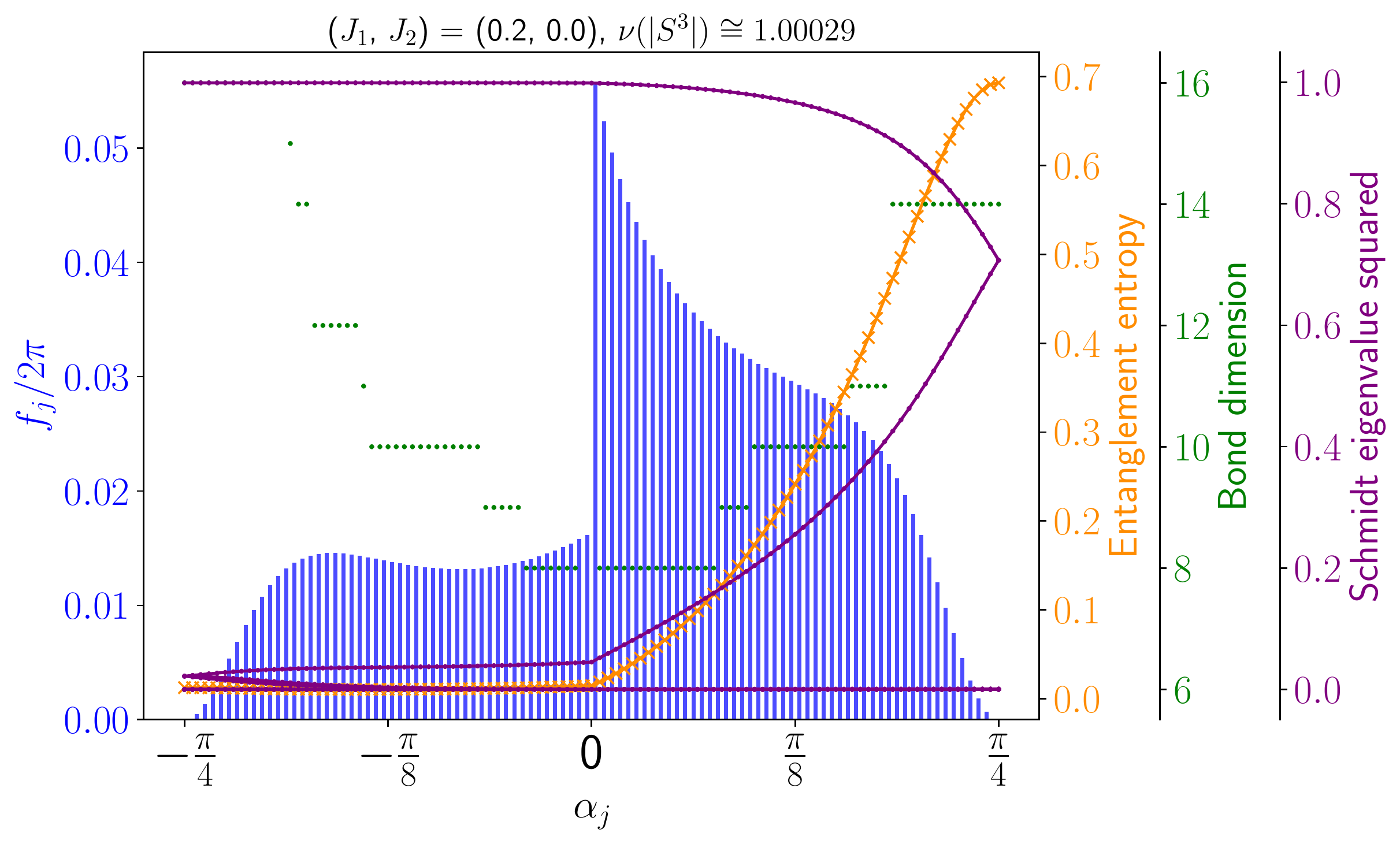} \\
&\includegraphics[width=0.5\linewidth, trim=0cm 0cm 0cm 0cm]{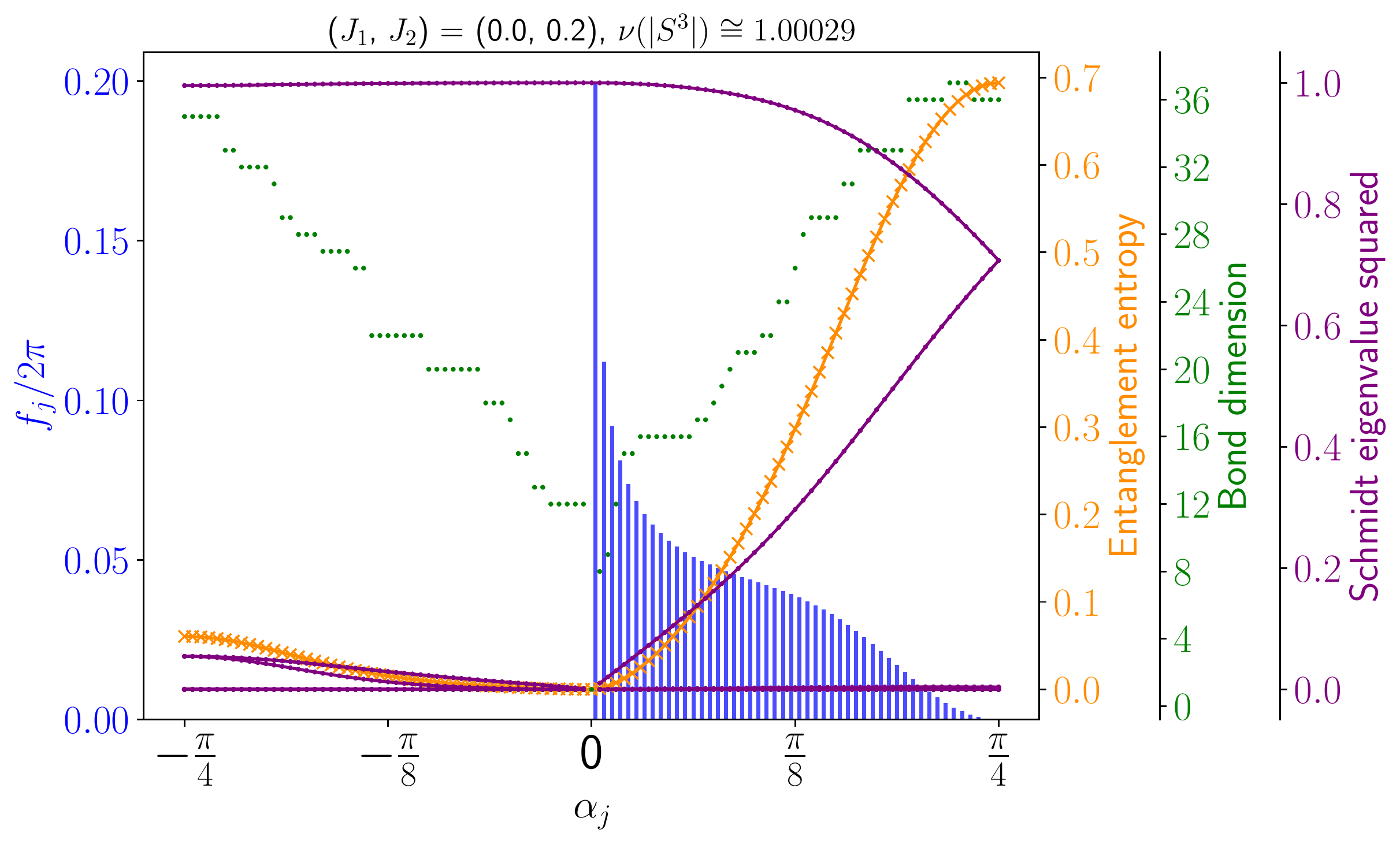} \quad & \quad 
\includegraphics[width=0.5\linewidth, trim=0cm 0cm 0cm 0cm]{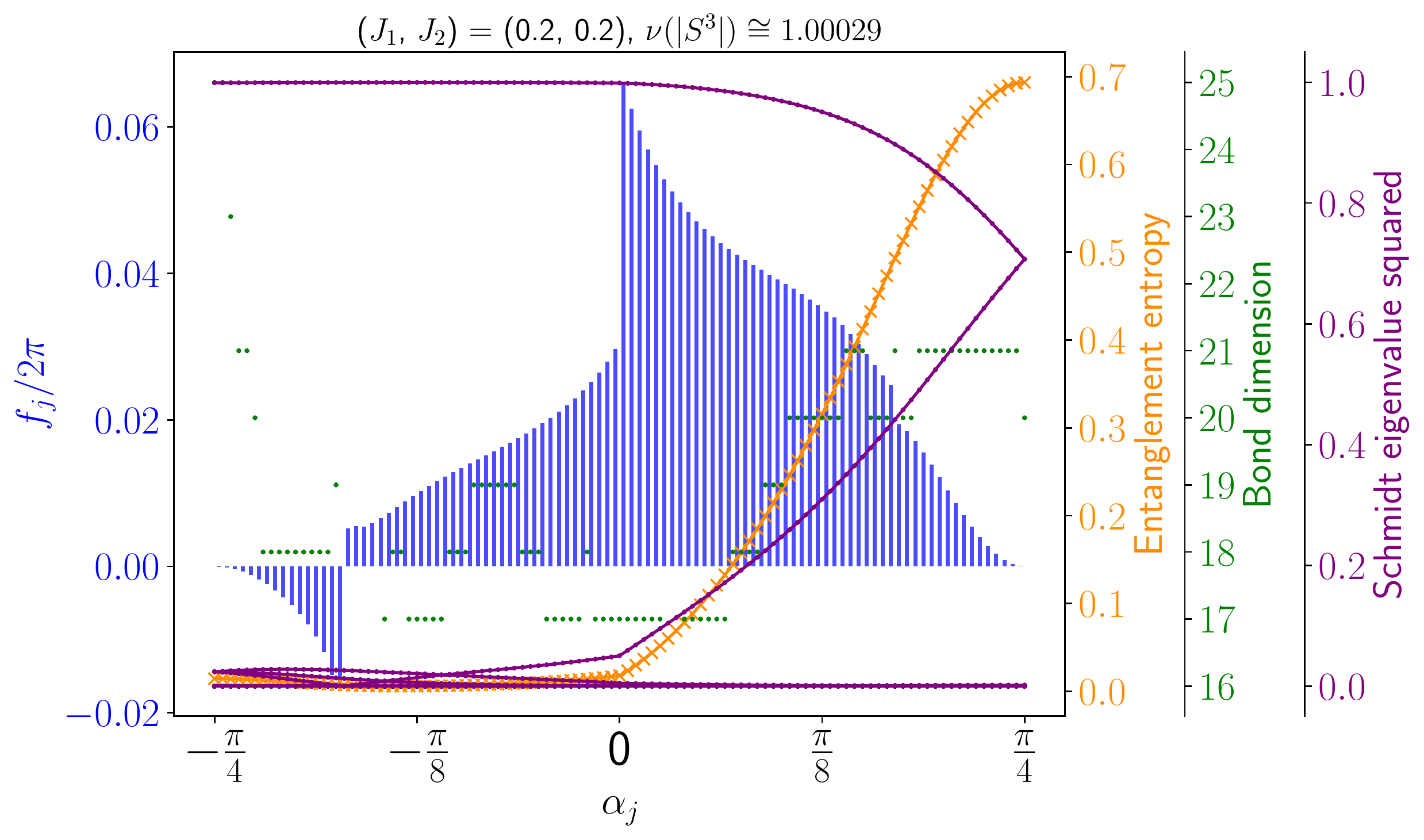} \\
\end{align*}
\caption{
Plots of the higher Berry curvature for $(J_1, J_2) = (0, 0), (0.2, 0), (0, 0.2), (0.2, 0.2)$. 
Blue bars represent the sum of the higher Berry curvature in the $S^2$ direction ($f_j/(2\pi)$ in eq.(\ref{eq:fj})), yellow and purple crosses indicate the entanglement entropy and the squares of Schmidt eigenvalues respectively, and green dots represent the bond dimension. 
The approximate value of the invariant $\nu(|S^3|)$ is shown at the top of the graph.
}
\label{fig:result}
\end{figure*}

Divide the interval $[-\frac{\pi}{4},\frac{\pi}{4}]$ into $N$ equal parts, and let $\alpha_j = -\frac{\pi}{4} + \frac{\pi j}{2 N}, j=0,\dots,N$. 
We calculate the MPS $\{A^i(\alpha_j,\hat z)\}_i$ of the Hamiltonian $H(\alpha_j,\hat z)$ using iDMRG as implemented in the TeNPy package~\cite{TeNPy}. 
Fix the point $(\theta_0,\phi_0)$ on $S^2$ and the small displacement vector $(\delta \theta,\delta \phi)$, and introduce a triangular prism $\Gamma$ with the following 6 vertices in each interval $[\alpha_j,\alpha_{j+1}]$:
\begin{align*}
&0: (\alpha_j,\theta_0-\frac{\delta \theta}{2},\phi_0-\frac{\delta \phi}{2}), \\
&1: (\alpha_{j+1},\theta_0-\frac{\delta \theta}{2},\phi_0-\frac{\delta \phi}{2}), \\
&2: (\alpha_j,\theta_0+\frac{\delta \theta}{2},\phi_0-\frac{\delta \phi}{2}), \\
&3: (\alpha_{j+1},\theta_0+\frac{\delta \theta}{2},\phi_0-\frac{\delta \phi}{2}), \\
&4: (\alpha_j,\theta_0-\frac{\delta \theta}{2},\phi_0+\frac{\delta \phi}{2}), \\
&5: (\alpha_{j+1},\theta_0-\frac{\delta \theta}{2},\phi_0+\frac{\delta \phi}{2}).
\end{align*}
Take the triangulation of the triangular prism $\Gamma$ as shown in Fig.~\ref{fig:cell} and orient it in the order of the numbers. 
The sum of the Berry curvatures in the triangular prism $\Gamma$ is given by $F(\Gamma) = F(0124)-F(1235)+F(1245)$.
The area of the small triangle $\delta S = (024)$ can be approximated as ${\rm Area}(\delta S)=\sin \theta_0 \delta \theta \delta \phi/2$, so the contribution of the higher Berry curvature from the spherical shell $[\alpha_j,\alpha_{j+1}]$ can be approximated as
\begin{align}
f_j = F(\Gamma) \times \frac{4\pi}{\sin \theta_0 \delta \theta \delta \phi/2}. 
\label{eq:fj}
\end{align}
Also, the total sum of the higher Berry curvatures is estimated as $\nu(|S^3|) = \sum_{j=0}^{N-1} f_j/(2\pi)$.

Fig.~\ref{fig:result} shows the plots of the sum of Berry curvatures $f_j/(2\pi)$ and the approximated total sum $\nu(|S^3|)$ for cases $(J_1, J_2) = (0, 0), (0.2, 0), (0, 0.2), (0.2, 0.2)$. 
We set $N = 100$, $(\theta_0, \phi_0) = (\pi/2, \pi)$, $\delta \theta = \pi/100$, and $\delta \phi = 2\pi/100$. 
Given the MPS $\{A^i(\alpha_j,\hat z)\}_i$, those at $(\theta,\phi)$ is computed by (\ref{eq:so3_unitary_tr}). 
In the iDMRG calculations, the singular value cutoff was set to $10^{-8}$, and the convergence criteria for the ground state energy and entanglement entropy were set to $\Delta E = 10^{-8}$ and $\Delta S = 10^{-8}$, respectively. 
The entanglement entropy and bond dimensions are also indicated in the figures. 
We confirmed that the topological invariant $\nu(|S^3|)$ takes the expected nontrivial quantized value of 1. 
(A small deviation from the exact value $1$ is because of the approximation (\ref{eq:fj}).)
The bond dimension depends on the truncation cutoff and convergence criteria in iDMRG, but the higher Berry curvature $f_j$ is insensitive to discontinuous changes in bond dimension. 
This indicates that the weighting by Schmidt eigenvalues $\Lambda^{\frac{2}{3}}$ in the higher Berry connection (\ref{eq:HBP_Delta^2}) is appropriately implemented. 
Fig.~\ref{fig:result} also shows the squares of Schmidt eigenvalues. 
For values of $\alpha$ less than zero, only a single Schmidt eigenvector predominantly contributes to the higher Berry curvature. However, when $\alpha$ is greater than zero, the contribution from the second Schmidt eigenvalue becomes significant. 
Moreover, in all cases, a discontinuous jump in the higher Berry curvature $f_j$ is observed at $\alpha = 0$. 
This is inferred from the fact that the model $H( \alpha, \bm{n})$ is not smooth with $\alpha = 0$.
This behavior is also common in the results by \cite{Xueda} at the non-perturbed parameter point $(J_1, J_2) = (0, 0)$ for the different formulation of the higher Berry curvature in \cite{KapustinSpodyneiko_Berry_curvature20}. 
In the non-perturbed Hamiltonian $(J_1, J_2) = (0, 0)$, the ground state is a tensor product state with zero higher Berry curvature in the range $\alpha \in [-\frac{\pi}{4}, 0]$. 
When the perturbation of $J_1$ is added, a finite higher Berry curvature arises even though the entanglement entropy is small. 
However, for the $J_2$ perturbation, the higher Berry curvature remained zero, despite small changes in bond dimension and entanglement entropy.

\subsubsection{Random $(J_1,J_2)$}
\begin{figure}[!]
\centering
\includegraphics[width=\linewidth]{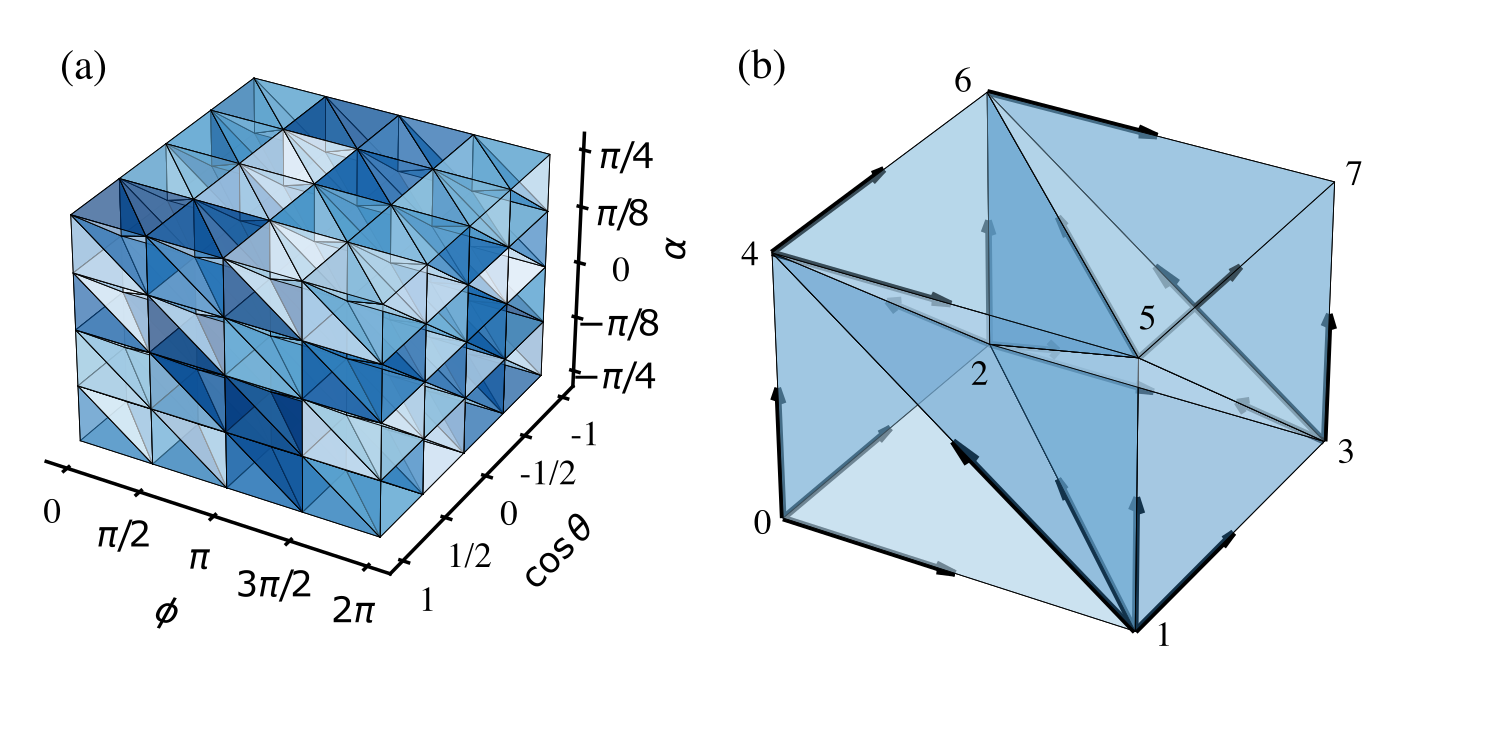}
\caption{
(a) A triangulation of the three-dimensional parameter space $|S^3|$. (b) A small rectangular prism constructed from 6 tetrahedra $\Delta^3$. The labels of its eight vertices are given and the orientation of the tetrahedron is indicated by the arrows.
}
\label{fig:cell_2}
\end{figure}

\begin{figure*}
    \centering
    \includegraphics{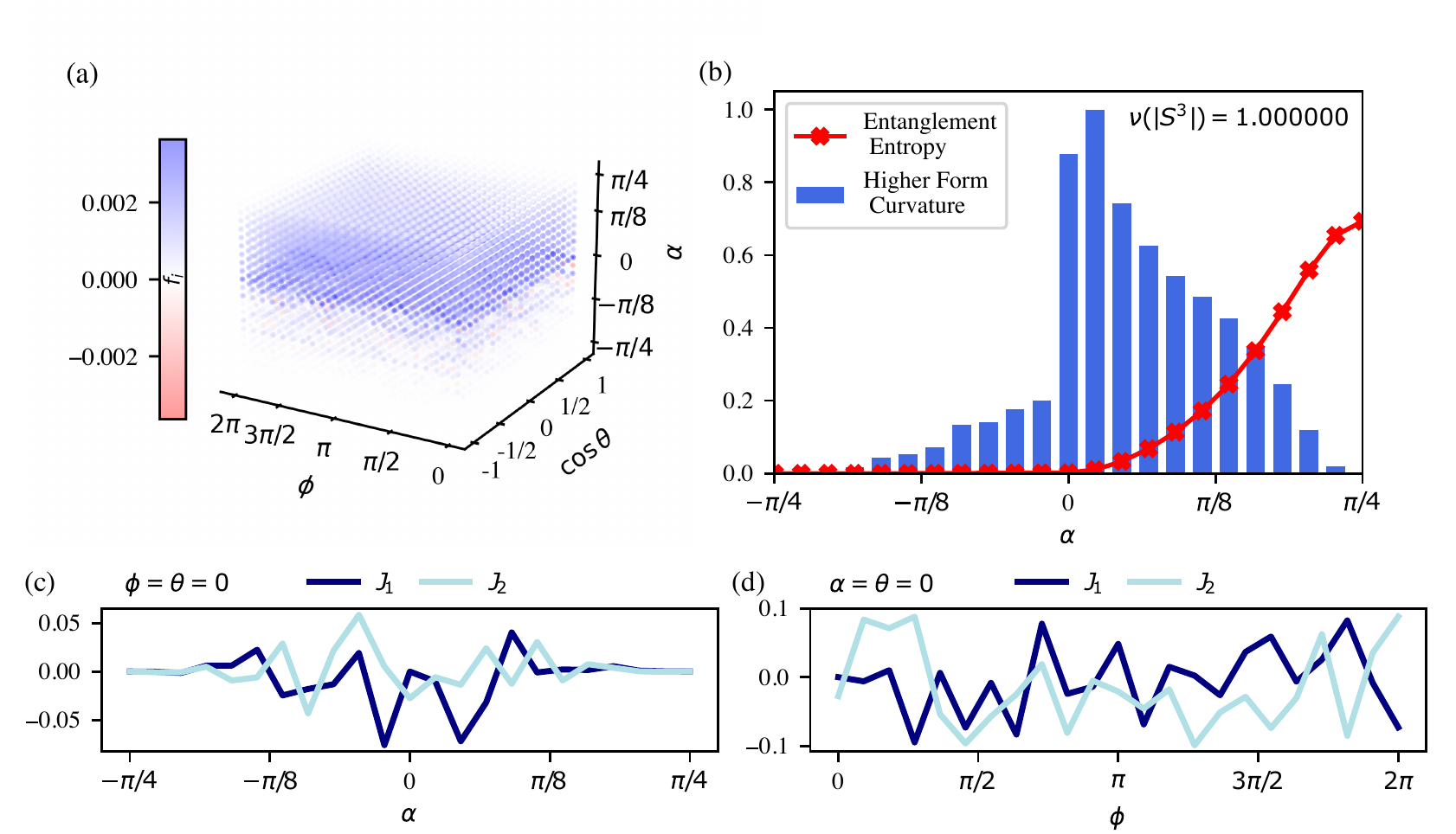}
    \caption{(a) Higher Berry curvature $f_i$ per rectangular prism as in Fig.\ \ref{fig:cell_2}(b) over the three-dimensional parameter space on a $23\times23\times23$ grid. (b) Blue bars represent the sum of higher Berry curvature in the $S^2$ direction ($f_j$ in eq.(\ref{eq:fj})), and the red crosses show the mean entanglement entropy. The numerical value of the invariant $\nu(|S^3|)$ is shown in the top right hand corner. (c) Perturbations $J_1$ and $J_2$ along the $S^2$ direction for $\phi = \theta = 0$. (d) Same as (c) but for constant $\alpha=\theta=0$ running over $\phi$.}
    \label{fig:result_2}
\end{figure*}

Fig.~\ref{fig:result_2} (a) shows higher Berry curvature $f_j$ and its sum $\nu(|S^3|)$ over the three-dimensional parameter space on a $23\times23\times23$ grid for random values of perturbations with $J_1, J_2 \in (-0.1, 0.1)$ as given in Fig.~\ref{fig:result_2} (c) and (d).
To maintain the parameter space as \( S^3 \), \( J_1 \) and \( J_2 \) are modulated by a Gaussian weight, ensuring they converge to zero at \( \alpha=\pm \pi/4 \).
Unlike the case where $J_1$ and $J_2$ are constant, the MPS must be computed at all grid points in the parameter space $(\alpha,\theta,\phi)$.
The parameter space is divided into cubes, and each cube is further divided into 6 tetrahedra, as shown in Fig.~\ref{fig:cell_2}.
In the iDMRG calculations, the singular value cutoff was set to $10^{-8}$, and the convergence criteria for the ground state energy and entanglement entropy were set to $\Delta E = 10^{-8}$ and $\Delta S = 10^{-8}$, respectively. The maximal bond dimension was set to $256$.
Fig.~\ref{fig:result_2} (b) shows the sums of higher Berry curvature in the $S^2$ direction as well as the mean entanglement entropy. We confirm that the topological invariant $\nu(|S^3|)$ takes the expected nontrivial quantized value of 1.

In all calculation results, the higher Berry curvature changes smoothly in the vicinity of the parameter space where the model is also smoothly defined, and the total sum over the entire $S^3$ is quantized to the expected value of 1, strongly suggesting that the formulation of higher Berry curvature in Sec.~\ref{sec:Higher Berry curvature} does work for more general one-dimensional Hamiltonians with translational symmetry.

\section{Conclusion}
\label{sec:conc}
In this paper, we discussed a formulation of the higher Berry curvature in one-dimensional quantum many-body systems using translationally invariant MPS. 
Based on the fact that a $U(1)$ function is defined in the triple patch intersection due to the gauge freedom of MPS~\cite{OhyamaRyu2023higher}, we introduced a discrete approximation of the higher Berry connection (\ref{eq:HBP_Delta^2}) using the eigenvectors of the mixed transfer matrices for edges of the discretized parameter space. 
Thanks to the weighting by Schmidt eigenvalues, the higher Berry connection (\ref{eq:HBP_Delta^2}) is free to the bond dimension. 
We defined the higher Berry phase (\ref{eq:HBP}) and used it to define the higher Berry curvature (\ref{eq:HBC}) in a small tetrahedron, like the lattice gauge theory. 
In Sec.~\ref{sec:Example of Numerical Calculation}, we showed numerical results of higher Berry curvature for a model, the unperturbed variant of which is known to have a non-trivial value~\cite{Xueda} of higher Berry curvature as proposed by Kapustin and Spodyneiko~\cite{KapustinSpodyneiko_Berry_curvature20}. 
We demonstrated that the higher Berry curvature we formulated varies smoothly in regions where the model parameter dependence is smooth, and the integral value in the three-dimensional parameter space is quantized to 1. 
These results suggest that the higher Berry curvature introduced in Sec.~\ref{sec:Higher Berry curvature} is expected to work in more general one-dimensional quantum many-body systems. 
It is important to compute the higher Berry curvature for a wider range of models to advance the understanding of the geometric structure of gapped states in quantum many-body systems.

\begin{acknowledgments}
KS thanks Kenji Shimomura and Yosuke Kubota for the useful discussion.
We especially thank Ryohei Kobayashi for his valuable suggestion to use MPS for formulating the higher Berry curvature at the beginning of this project.
KS was supported by JST CREST Grant No. JPMJCR19T2, and JSPS KAKENHI Grant No. 22H05118 and 23H01097. N.H. acknowledges support by the Deutsche Forschungsgemeinschaft (DFG, German Research Foundation) – TRR 360 – 492547816.
S.O. was supported by the establishment of university fellowships towards the creation of science technology innovation.
We thank the YITP workshop YITP-T22-02 on ``Novel Quantum States in Condensed Matter 2022", which was useful in completing this work.
\end{acknowledgments}

\appendix

\section{Mixed transfer matrix between different SPT phases}
\label{app:Gsym_transfer_mat}
Let $G$ be a finite group and $\phi: G \to \{\pm 1\}$ be a homomorphism so that $\phi_g = 1$ represents unitary symmetry, while $\phi_g = -1$ represents anti-unitary symmetry. 
For a matrix $X$, we introduce the notation
\begin{align}
X^{\phi_g} = \left\{\begin{array}{ll}
X & (\phi_g=1), \\
X^* & (\phi_g=-1). \\
\end{array}\right.
\end{align}
Let $\{\ket{i_l}\}_{i_l=1}^d$ be the local basis at site $l \in \Z$. 
Let $\{u_g\}_{g \in G}$ be a $d$-dimensional linear unitary representation of $G$. 
It satisfies $u_g u_h^{\phi_g} = u_{gh}$. 
For the Hilbert space $\bigotimes_{l=1}^L \bigoplus_{i_l=1}^d \C[\ket{i_l}]$, we introduce the action of $G$ as 
\begin{align}   
\hat g = \bigotimes_{l=1}^L \hat u^{[l]}_g,\quad 
\hat u^{[l]}_g \ket{j_l} = \sum_{i_l=1}^d \ket{i_l} [u_g]_{i_lj_l}. 
\end{align}

Let $\{A^i\}_i$ be an injective and canonical $D$-MPS. 
We denote the positive-definite diagonal matrix in the left-canonical condition (\ref{eq:left_cano}) by $\Lambda$. The action of $G$ on the MPS $\ket{\{A^i\}_i}_L$ is given by
\begin{align}
    \hat g \ket{\{A^i\}_i}_L
    = \Ket{\left\{\sum_{j=1}^d [u_g]_{ij}(A^j)^{\phi_g}\right\}_i}_L. 
\end{align}
Thus, according to the fundamental theorem of MPS~\cite{MPS}, when the MPS $\ket{\{A^i\}_i}$ has $G$ symmetry, there exists $e^{i\theta_g} \in U(1)$ and $V_g \in U(D)$ such that
\begin{align}
&\sum_{j=1}^d [u_g]_{ij} [A^j]^{\phi_G} = e^{i\theta_g} V_g^\dag A^i V_g, \\
&[V_g, \Lambda] = 0, 
\end{align}
hold. 
(Here, we used the fact that $\Lambda$ is invariant because the left-canonical condition (\ref{eq:left_cano}) is invariant under the transformation $\{A^i\}_i \to \{[u_g]_{ij}(A^j)^{\phi_g}\}_i$.) 
$e^{i\theta_g}$ is unique, and $V_g$ is unique up to a $U(1)$ phase. 
Moreover, by an argument similar to that in Sec.~\ref{sec:Cocycle condition}, $e^{i\theta_g}$ is a linear one-dimensional representation of $G$, i.e.,
\begin{align}
e^{i\theta_g} e^{i \phi_g \theta_h} = e^{i\theta_{gh}},\quad g,h \in G, 
\end{align}
holds, and there exists a 2-cocycle $\omega_{g,h} \in Z^2(G,U(1)_\phi)$ such that
\begin{align}
V_g V_h^{\phi_g} = \omega_{g,h} V_{gh},\quad g,h \in G, 
\end{align}
which means that $V_g$ is an $\omega$-projective representation. 
$G$-symmetric and translationally symmetric MPS are classified by $(e^{i\theta_g}, [\omega_{g,h}]) \in H^1(G,U(1)_\phi) \times H^2(G,U(1)_\phi)$~\cite{PhysRevB.85.075125,PhysRevB.83.035107,PhysRevB.84.165139}.

Now, let $\{A_0^i\}_i$ and $\{A_1^i\}_i$ be $G$-symmetric $D_0$- and $D_1$-MPSs belonging to SPT phases classified by $(e^{i\theta^0_g},[\omega^0_{g,h}])$ and $(e^{i\theta^1_g},[\omega^1_{g,h}])$, respectively.
Introduce the mixed transfer matrix $T_{01}$ by (\ref{eq:minxed_TM}). 
Fix one 2-cocycle $\omega^n_{g,h}$ for each $n=0,1$, and denote the $\omega^n$-projective representation by $V^n_g$. 
Noting that
\begin{align}
&A_n^i = \left(\sum_j [u_g^\dag]_{ij} e^{i\theta^n_g} (V^n_g)^{\dag} A_n^j V^n_g\right)^{\phi_g}, 
\end{align}
a straightforward calculation yields the $G$ symmetry of the mixed transfer matrix $T_{01}$:
\begin{align}
&V^{0}_g T_{01}(X)^{\phi_g} (V^1_g)^\dag\nonumber\\
&=e^{i(\theta^0_g-\theta^1_g)} 
T_{01}\left(V^0_g X^{\phi_g} (V^{1}_g)^{\dag}\right)
\end{align}
Introducing the matrix element representation as
\begin{align}
    &[T_{01}]_{ab,cd}=\sum_{i=1}^d [A_0^i]_{ac} [A_1^{i*}]_{bd}, \\
    &[V_g \otimes V_g^*]_{ab,cd} = [V_g]_{ac} [V_g^*]_{bd}
\end{align}
we obtain the matrix representation of the $G$ symmetry:
\begin{align}
&(V^0_g \otimes V^{1*}_g) [T_{01}]^{\phi_g}\nonumber\\
&=e^{i(\theta^0_g-\theta^1_g)} T_{01} (V^0_g \otimes V^{1*}_g). 
\label{eq:Gsym_T01}
\end{align}

We now prove that if two MPSs belong to different SPT phases, i.e., $(e^{i\theta^0_g},[\omega^0_{g,h}])\neq (e^{i\theta^1_g},[\omega^1_{g,h}])$, the absolute value of the eigenvalue $\eta$ of the mixed transfer matrix $T_{01}$ must be degenerate. 
Let $\ket{\eta}$ be the right eigenvector of $T_{01}$ with eigenvalue $\eta$. 
If $e^{i\theta_g^0}$ and $e^{i\theta_g^1}$ are different one-dimensional representations, there exists some $g \in G, \phi_g=1$ such that $e^{i(\theta^0_g-\theta^1_g)} \neq 1$, and for such $g\in G$, $e^{-i(\theta_g^0-\theta_g^1)} \eta$ is also an eigenvalue. 
On the other hand, if $e^{i\theta^0_g}$ and $e^{i\theta^1_g}$ are the same one-dimensional representation of $G$, the transfer matrix $T_{01}$ satisfies the symmetry
\begin{align}
(V^0_g \otimes V^{1*}_g) [T_{01}]^{\phi_g}
=T_{01} (V^0_g \otimes V^{1*}_g).
\end{align}
The representation $V^0 \otimes V^{1*}$ is an $\omega^0 \omega^{1*}$-projective representation, and if $[\omega^0] \neq [\omega^1]$, $[\omega^0 \omega^{1*}] \in H^2(G,U(1)_\phi)$ is a nontrivial element. 
Therefore, the dimension of the irreducible representation is at least 2, and if $\eta$ is an eigenvalue of $T_{01}$, then either $\eta$ or $\eta^*$ is also an eigenvalue.

\section{MPS of Hamiltonian (\ref{eq:H0})}
\label{app:MPS_Xueda}
For $\alpha \in [-\pi/4,0]$, the ground state of the Hamiltonian $H_0(\alpha,\hat z)$ is 
\begin{align}
    &\ket{GS(\alpha,\hat z)} \nonumber\\
    &= \bigotimes_{j\in \Z} \left( \cos \alpha \ket{\ua}_{\s_j}\ket{\da}_{\tau_j} + \sin \alpha \ket{\da}_{\sigma_j} \ket{\ua}_{\tau_j} \right), 
\end{align}
and MPS is 
\begin{align}
    &(A^{\ua\ua}(\alpha,\hat z),A^{\ua\da}(\alpha,\hat z),A^{\da\ua}(\alpha,\hat z),A^{\da\da}(\alpha,\hat z))\nonumber\\
    &=(0,\cos \alpha,\sin \alpha,0). 
\end{align}
For $\alpha \in [0,\frac{\pi}{4}]$, 
\begin{align}
    &\ket{GS(\alpha,\hat z)}\nonumber\\
    &=\bigotimes_{j \in \Z} (\cos \alpha \ket{\ua}_{\tau_j} \ket{\da}_{\s_{j+1}} - \sin \alpha \ket{\da}_{\tau_j} \ket{\ua}_{\s_{j+1}} ), 
\end{align}
and an example of MPS is 
\begin{align}
    &(A^{\ua\ua}(\alpha,\hat z),A^{\ua\da}(\alpha,\hat z),A^{\da\ua}(\alpha,\hat z),A^{\da\da}(\alpha,\hat z))\nonumber\\
    &\left( \begin{pmatrix}
0&0\\
-\sin \alpha&0\\
\end{pmatrix},\begin{pmatrix}
0&0\\
0&\cos \alpha\\
\end{pmatrix},\begin{pmatrix}
-\sin \alpha&0\\
0&0\\
\end{pmatrix}, \begin{pmatrix}
0&\cos \alpha\\
0&0\\
\end{pmatrix} \right).
\end{align}
The MPS for generic $\bm{n} \in S^2$ is given by the unitary transformation (\ref{eq:utr_for_S2}) 
\begin{align}
    A^{\s\tau}(\alpha,\bm{n})
    =\sum_{\s'\tau'} [u(\bm{n})]_{\s\tau,\s'\tau'} A^{\s'\tau'}(\alpha,\hat z). 
\end{align}
Explicitly, 
\begin{align}
&(A^{\ua\ua}(\alpha,\bm{n}),A^{\ua\da}(\alpha,\bm{n}),A^{\da\ua}(\alpha,\bm{n}),A^{\da\da}(\alpha,\bm{n}))\nonumber\\
&=
\begin{pmatrix}
-\frac{1}{2} e^{-i \phi } (\cos (\alpha )+\sin (\alpha )) \sin (\theta )\\
\cos (\alpha ) \cos ^2\left(\frac{\theta }{2}\right)-\sin (\alpha ) \sin ^2\left(\frac{\theta }{2}\right)\\
\cos ^2\left(\frac{\theta }{2}\right) \sin (\alpha )-\cos (\alpha ) \sin ^2\left(\frac{\theta }{2}\right)\\
\frac{1}{2} e^{i \phi } (\cos (\alpha )+\sin (\alpha )) \sin (\theta ) 
\end{pmatrix}^T
\end{align}
for $\alpha \in [-\pi/4,0]$, and 
\begin{align}
&(A^{\ua\ua}(\alpha,\bm{n}),A^{\ua\da}(\alpha,\bm{n}),A^{\da\ua}(\alpha,\bm{n}),A^{\da\da}(\alpha,\bm{n}))\nonumber\\
&=
\begin{pmatrix}
\begin{pmatrix}
    \frac{1}{2} e^{-i \phi } \sin (\alpha ) \sin (\theta )&e^{-i \phi } \cos (\alpha ) \sin ^2\left(\frac{\theta }{2}\right)\\
    -e^{-i \phi } \cos ^2\left(\frac{\theta }{2}\right) \sin (\alpha )&-\frac{1}{2} e^{-i \phi } \cos (\alpha ) \sin (\theta ) \\
\end{pmatrix}\\
\begin{pmatrix}
 \sin (\alpha ) \sin ^2\left(\frac{\theta }{2}\right)&-\frac{1}{2} \cos (\alpha ) \sin (\theta ) \\
 -\frac{1}{2} \sin (\alpha ) \sin (\theta )&\cos (\alpha ) \cos ^2\left(\frac{\theta }{2}\right) \\
\end{pmatrix}\\
\begin{pmatrix}
 -\cos ^2\left(\frac{\theta }{2}\right) \sin (\alpha )&-\frac{1}{2} \cos (\alpha ) \sin (\theta ) \\
 -\frac{1}{2} \sin (\alpha ) \sin (\theta )&-\cos (\alpha ) \sin ^2\left(\frac{\theta }{2}\right) \\
\end{pmatrix}\\
\begin{pmatrix}
-\frac{1}{2} e^{i \phi } \sin (\alpha ) \sin (\theta )&e^{i \phi } \cos (\alpha ) \cos ^2\left(\frac{\theta }{2}\right) \\
-e^{i \phi } \sin (\alpha ) \sin ^2\left(\frac{\theta }{2}\right)&\frac{1}{2} e^{i \phi } \cos (\alpha ) \sin (\theta ) \\
\end{pmatrix}\\
\end{pmatrix}^T
\label{eq:MPS_Xueda_D2}
\end{align}
for $\alpha \in [0,\pi/4]$.
Here, $\bm{n} = (\theta,\phi)$ is the polar coordinate.
At $\theta=0$, the MPS in (\ref{eq:MPS_Xueda_D2}) still exhibits $\phi$-dependence, implying an ill-defined gauge choice at this point. However, despite this, the formulation of higher Berry curvature outlined in Sec.~\ref{sec:Higher Berry curvature} remains applicable to the MPS (\ref{eq:MPS_Xueda_D2}) due to its gauge independence.

\bibliography{ref}

\end{document}